\documentclass{cta-author}

\usepackage{lipsum}  
\usepackage{booktabs}

\usepackage{siunitx}
\DeclareSIUnit{\pu}{p.u.}
\usepackage{subcaption}
\usepackage{amsmath}

\begin{document}

\supertitle{Modelling of Variable-Speed Refrigeration for Fast-Frequency Control in Low-Inertia Systems}

\title{Modelling of Variable-Speed Refrigeration for Fast-Frequency Control in Low-Inertia Systems}

\author{\au{Johanna Vorwerk$^{1^\star}$}, \au{Uros Markovic$^{1}$}, \au{Petros Aristidou$^{2}$}, \au{Evangelos Vrettos$^3$}, \au{Gabriela Hug$^1$}}

\address{\add{1}{Power Systems Laboratory, ETH Zürich, Physikstrasse 3, 8092 Zürich, Switzerland}
\add{2}{Department of Electrical Engineering, Computer Engineering and Informatics, Cyprus University of Technology}
\add{3}{Swissgrid AG, was with Grid Integration Group at Lawrence Berkeley National Lab while main body of the work was performed}
\email{vorwerk@eeh.ee.ethz.ch}}

\begin{abstract}
In modern power systems, shiftable loads contribute to the flexibility needed to increase robustness and ensure security. Thermal loads are among the most promising candidates for providing such service due to the large thermal storage time constants. This paper demonstrates the use of Variable-Speed Refrigeration (VSR) technology, based on brushless DC motors, for fast-frequency response. First, we derive a detailed dynamic model of a single-phase VSR unit suitable for time-domain and small-signal stability analysis in low-inertia systems. For analysing dynamic interactions with the grid, we consider the aggregated response of multiple devices. However, the high computational cost involved in analysing large-scale systems leads to the need for reduced-order models. Thus, a set of reduced-order models is derived though transfer function fitting of data obtained from time-domain simulations of the detailed model. The modelling requirements and the accuracy versus computational complexity trade-off are discussed. Finally, the time-domain performance and frequency-domain analyses reveal substantial equivalence between the full- and suitable reduced-order models, allowing the application of simplified models in large-scale system studies.
\end{abstract}

\maketitle

\section{Introduction}

In recent decades, the increasing share of renewable generation poses new challenges for electricity grids. Renewable Energy Sources (RESs) are typically interfaced asynchronously to the grid through power electronic converters and therefore do not contribute to system inertia or damping. As a result, faster frequency dynamics occur, thus increasing the need for rapid frequency regulation~\cite{Milano18, Fang19}. Consequently, new resources are required to provide power balancing services on different timescales. Moreover, converter-interfaced generation and load units introduce a distinct timescale separation, characteristic of low-inertia systems, which might disrupt frequency and voltage stability~\cite{markovic_2019}. Thus, to conduct stability studies of low-inertia grids, detailed and accurate models in the form of Differential-Algebraic Equations (DAEs) are needed~\cite{Milano18}. 

Traditionally, frequency reserves were provided by large conventional generation units. However, the activation time requirement for new rapid reserve provision services is very short and can predominantly be achieved by converter-interfaced units. In addition to non-synchronous generation, such as wind and photovoltaic power plants, thermal loads are among the most prominent candidates. While the provision of fast reserve with RES generation units usually comes at the expense of energy curtailment, thermal loads can simply shift their energy consumption due to inherent thermal inertia. Several transmission system operators (e.g., EirGrid in Ireland and National Grid in Great Britain) have already established Fast Frequency Reserve (FFR) provision programs to support system stability in times of low inertia and allow for demand-response contribution~\cite{Fernandez20}. To date, \SI{20}{\percent} of FFR in Ireland are provided by demand-side units~\cite{Fernandez20}.

The existing literature proposes several control schemes for FFR provision by thermal loads. In particular, \cite{ziras_primary_2015} presents a probabilistic switching scheme that allows a large population of residential refrigerators to provide primary frequency response using only frequency measurements. However, achieving a coordinated response within such a population of conventional devices, that can only operate at a fixed power or be turned off, is difficult as it requires considering and estimating the behaviour of all contributing units. In contrast, emerging Variable-Speed Drive (VSD) technologies, studied in \cite{kim_modeling_2015,hui_equivalent_2019,chakravorty_rapid_2017,Che19,Azizipanah20, Malekpour20,Ibrahim20}, offer a continuous controllable range for a single unit, hence simplifying the aggregated control. 

The application of VSD heat pumps based on induction motors is analysed in \cite{kim_modeling_2015}, whereas the studies in \cite{hui_equivalent_2019,Che19} consider inverter-interfaced air conditioning. In general, the timescales of the considered applications lie in the range of primary frequency control schemes, without considering the inertial response. In addition, the modelling of the VSD and motor considered in these studies is simplistic and employs reduction of certain system dynamics without verification of the validity of such reduction. More precisely, the entire load unit is typically subsumed within a single low-order transfer function and the dynamic interactions between the units and the grid are usually neglected. 

Some of the aforementioned drawbacks are addressed in \cite{chakravorty_rapid_2017, Azizipanah20, Malekpour20, Ibrahim20}. The authors in \cite{chakravorty_rapid_2017} estimate the potential of industrial induction motors in Great Britain for providing FFR using local droop control. The size of the considered reserve is comparable to spinning reserve in Great Britain, but ramp-rates are introduced to avoid regeneration of unidirectional units. This approach has been extended in \cite{Azizipanah20, Malekpour20}, which designs a droop-based control that replicates the inertial response of an online motor, while \cite{Malekpour20} develops an adaptive control scheme to avoid regeneration of the induction motor. Nonetheless, only the secondary-side dynamics of the drives are included and the inertia present in the considered test cases is not particularly low. Furthermore, \cite{Malekpour20} showcases a reduced efficacy of the control scheme for a lower inertia scenario. In contrast, \cite{Ibrahim20} presents a small-signal model of higher order for a VSD, induction-machine-based heat pump that can provide inertial response and FFR. The model is verified through hardware-in-the-loop simulations, and time-domain simulations are performed for a sample distribution system. Still, some of the fast dynamics are neglected and stability analysis is not performed. 

VSD applications are not only emerging for heat pumps and induction motors, but are increasingly used in refrigeration to cope with recent energy efficiency standards \cite{BFE, FEA18}. These refrigeration systems typically contain a Brushless DC Motor (BLDC) and a variable-speed compressor. Their contribution to FFR has received little attention in the literature, even though the overall potential could be significant. This is supported by the studies in \cite{Gils14,Grein06}, that evaluate the theoretical Demand Response (DR) potential in Europe for various load types and industries \cite{Gils14}, and estimate the accessible volumes for refrigeration devices~\cite{Grein06}. The results suggest a theoretical volume of refrigeration and freezers of \SI{2.8}{\giga \watt} in Germany and \SI{15}{\giga\watt} in Europe. For comparison, the latter is five-fold the volume of the current primary spinning reserve in the power grid of continental Europe~\cite{UCTE}. In contrast, the capability for DR contribution of heat pumps is half of the one estimated for refrigeration~\cite{Gils14}. While heat pumps inherit a higher power rating per device, they face a substantial disadvantage as the reserve is significantly reduced when outdoor temperatures are changing. As a result, they hinder the forecasting of their potential contribution. On the contrary, refrigeration units usually locate indoors, and therefore offer a less fluctuating reserve availability.

\begin{figure*}[b]
    \centering
    \scalebox{0.8}{\includegraphics{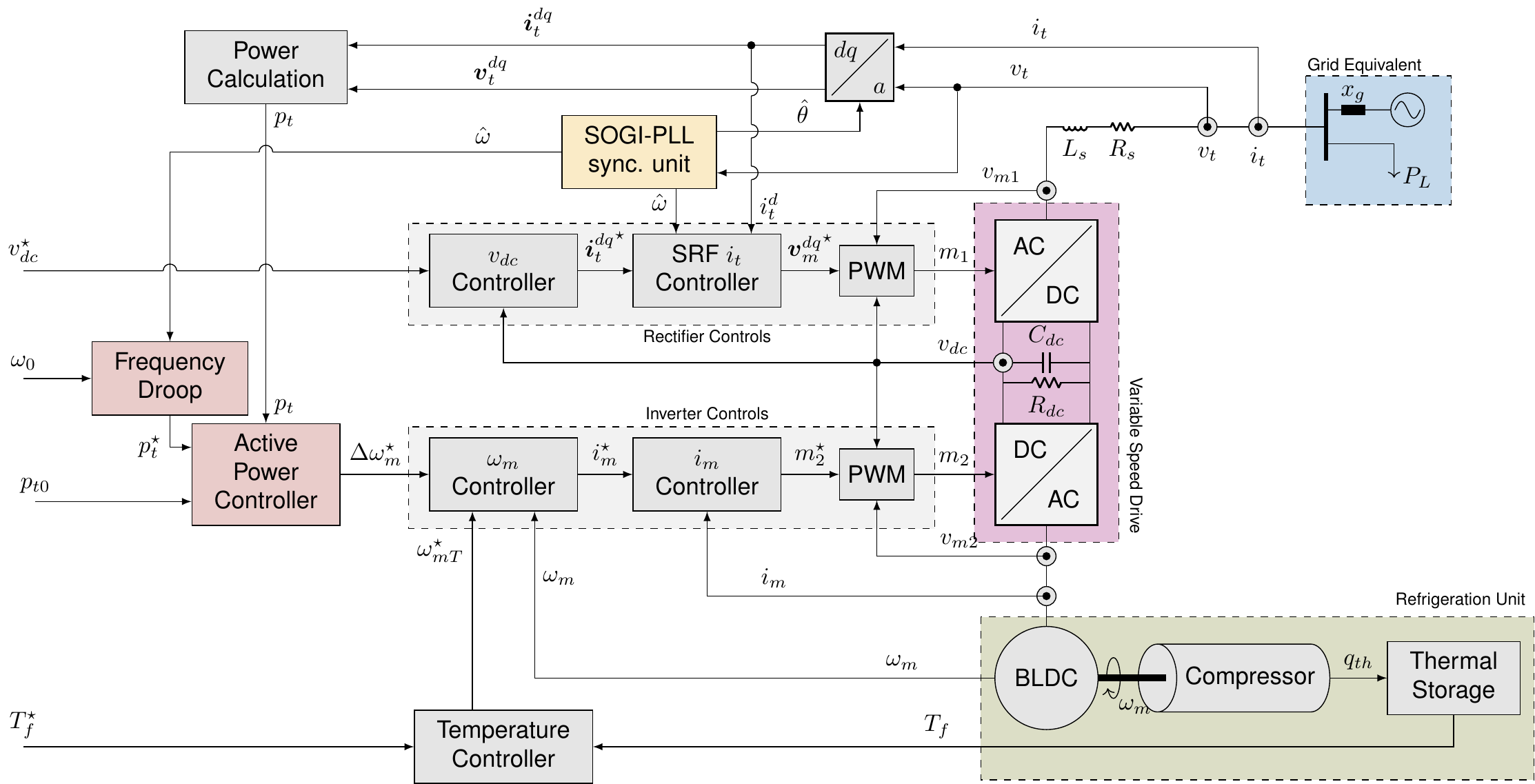}}
    \caption{Investigated system configuration and VSDR control structure.}
    \label{fig:ModelOverview}
\end{figure*}

In this paper, we analyse the capability of BLDC-based VSD Refrigeration (VSDR) for providing FFR, while addressing the requirement for detailed DAE models in low-inertia system studies. We propose a decentralised Fast-Frequency Control (FFC) design for VSDR technology with several contributions. First, a detailed DAE model of BLDC-based VSD technology is derived. While only the basic thermal dynamics of the refrigerator are considered, as is common in power system studies, the dynamics of electric components are integrated in full detail, which allows for capturing the complex dynamic interactions and limitations of the units. Second, a BLDC-based FFC scheme for supply of synthetic inertia through VSDR is proposed and verified through time-domain simulations. In addition, we propose a process to decide which reduced-order models to apply for a given detailed model. Therefore, six reduced-order models that are potentially useful for large-scale studies and allow for efficient parameterisation are derived and compared to the original model in terms of time-domain performance. A further important contribution of this paper is the comparison of the small-signal stability analysis for the various models which is missing in related work.

The remainder of this paper is structured as follows. In Section~\ref{Sec2}, we introduce the detailed droop-based VSDR control scheme and the detailed DAE and Small-Signal Model (SSM) formulation. Section~\ref{sec3} proposes six reduced-order models and determines the respective parameters by utilising the time-domain simulations of the proposed detailed model. Subsequently, Section~\ref{sec4} demonstrates the time-domain performance of the full-order model and compares it to the suggested reduced models, whereas the stability analysis is performed in Section~\ref{sec5}. Finally, Section~\ref{sec6} discusses the outlook of the study and concludes the paper. 
\vspace{-10pt}
\section{VSDR Control Scheme} \label{Sec2}
An overview of the implemented control scheme is shown in Fig.~\ref{fig:ModelOverview}, where a VSDR unit is connected to a single-phase distribution grid. The VSDR comprises of a thermal storage and a variable capacity compressor. The latter is driven by a three-phase BLDC motor and connected to the grid via two back-to-back, full-bridge converters: a rectifier that maintains a constant DC-link voltage and an inverter that regulates the shaft frequency. The applied speed control coordinates two objectives: in the long term, it keeps the temperature of the refrigeration compartment within bounds, and in the short term, it supports the grid frequency by adjusting the shaft rotational speed and hence the active power consumption at the Point of Common Coupling (PCC). All parameters and variables used in Fig.~\ref{fig:ModelOverview} are defined in the subsequent sections.

Due to the trapezoidal shape of the BLDC back electromagnetic force (EMF), the modelling, analysis and control of the BLDC and inverter are implemented in a stationary frame. Despite the single-phase connection of the unit, the grid-side rectifier unit is controlled and modelled in a Synchronously-rotating Reference Frame (SRF) to allow for small-signal analysis. The complete mathematical model is defined in per-unit, denoted by lower-case symbols. All control setpoints, i.e., exogenous control inputs and internally computed reference signals, are represented by $x^\star$, estimated parameters are marked with $\hat{x}$, whereas signal magnitudes are described as $\overline{x}$. First and second-order derivatives are denoted by $\dot{x}$ and $\Ddot{x}$, respectively. Note that all integrator states are represented by $\mu_i$, with $i$ being the identification subscript.

The remainder of this section presents every system component and the corresponding model individually. Starting from the thermal dynamics in the refrigeration chamber and the variable capacity compressor, we continue with the BLDC and the variable-speed drive before completing the model with the outer control loop (i.e., the active power control), the speed reference computation, and the grid equivalent.

\subsection{Variable-Speed Refrigeration Unit}\label{sec:VSDRunit}

\begin{figure}[b]
\vspace{-5pt}
\begin{subfigure}[t]{0.48\linewidth}
    \scalebox{0.75}{\includegraphics{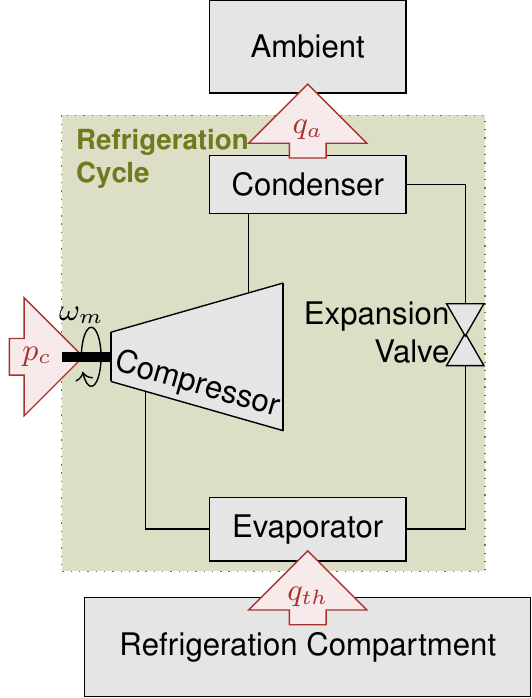}}
\end{subfigure}
    \hfill
\begin{subfigure}[t]{0.5\linewidth}
    \scalebox{1}{\includegraphics{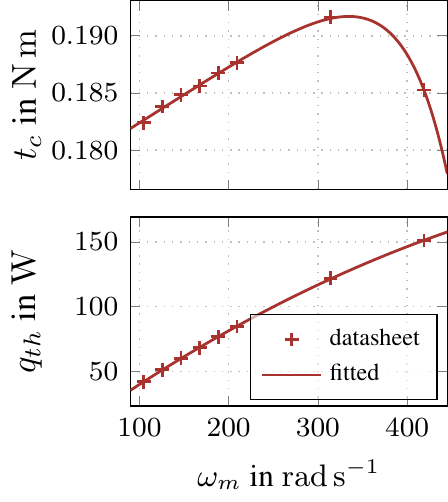}}
\end{subfigure}
\caption{Schematic diagram of the refrigeration cycle (left) and the steady-states compressor model (right).} \label{fig:Refrigerator}
\vspace{-5pt}
\end{figure}

The VSDR unit includes a refrigeration compartment with thermal load and thermal cycle that discharges heat $(q_{th})$ from the cold compartment and transfers it to the ambient air $(q_a)$, as shown in Fig.~\ref{fig:Refrigerator}. Several models for VSD vapour compression technologies, particularly focusing on VSD heat pumps and air conditioners, exist in the literature, with different modelling approaches and level of detail~\cite{koury_2001,kim_modeling_2015,hui_equivalent_2019,verhelst_2012}. In particular, \cite{koury_2001} presents two detailed numerical models based on partial differential equations to simulate steady-state and transient behaviour of the vapour compression cycle for VSDR units, whereas \cite{kim_modeling_2015,hui_equivalent_2019,verhelst_2012} exploit data-driven models that incorporate the slow dynamics of the entire thermal cycle within a first or second-order transfer function. For the scope of this paper, the power consumption of the compressor $(p_c)$ and the removed heat from the cold chamber are of primary interest. Therefore, the applied refrigeration model includes the compressor but neglects other system components, such as the evaporator and condenser. 
\begin{figure}[b]
    \centering
    \includegraphics{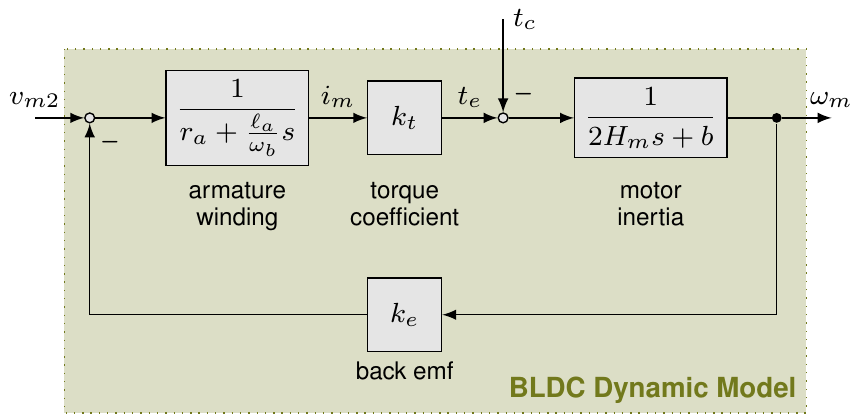}
    \caption{Block diagram of the BLDC model.}
    \label{fig:BLDC}
\end{figure}

\subsubsection{Variable-Speed Compressor}
Based on the dynamic model of VSD heat pumps in \cite{kim_modeling_2015}, the dynamics of the compressor torque $t_c$ and the heat removed $q_{th}$ can be described as
\begin{subequations}
\label{eq:qth_pc}
\vspace{-5pt}
\begin{align}
    \dot{q}_{th} &= \frac{1}{\tau_q}\, \bigl(\,\overbrace{a_2 \omega_m^2 + a_1 \omega_m + a_0}^{\displaystyle q_{th,0}}\,-\,q_{th}\,\bigr),  \label{eq:qth}\\
    \dot{t}_{c} &= \frac{1}{\tau_c}\, \bigl(\,\underbrace{b_1 e^{b_2 \omega_m} + b_3 e^{b_4 \omega_m}}_{\displaystyle t_{c0}}\,-\,t_{c}\,\bigr). \label{eq:pc}
\end{align}
\end{subequations}
Despite the similarities in thermal cycles of heat pumps and refrigerators, the dependency of the two modelled quantities on the ambient and evaporator temperature is neglected in this case and both variables are assumed to be constant. This is justified as the indoor placement limits the variations in the former temperature, while the superheat control maintains the latter \cite{li_feedforward_2009}. While the steady-state response (denoted by $q_{th,0}$ and $t_{c,0}$) is a non-linear function of the shaft speed, the transient response is described by first-order transfer functions. All coefficients ($a_i,b_i$) in \eqref{eq:qth_pc} have been obtained through curve fitting of SECOP compressor data sheets \cite{Secop_Data}, whereas the time delays ($\tau_q, \tau_c$) have been determined using additional data provided by the manufacturers. The accuracy of the steady-state compressor model fitting is showcased in Fig.~\ref{fig:Refrigerator}. 

\subsubsection{Refrigeration Compartment}
Similar to the approach taken in \cite{ziras_primary_2015}, the refrigeration compartment is modelled with a first-order differential equation. Hence, a unique compartment temperature $T_f$ is assumed, that depends on thermal resistance $r_{th}$, thermal capacitance $c_{th}$ and supplied heat $q_{th}$ as follows:
\begin{align}
\vspace{-5pt}
    \dot{T}_f &= \frac{T_a-T_f}{r_{th} c_{th}} - \frac{q_{th}}{c_{th}},
    \vspace{-5pt}
\end{align}
with the ambient temperature $T_a$ assumed to be constant.

\subsection{Brushless DC Motor}
The torque required by the compressor is supplied by a BLDC. Compared to induction machines, BLDCs offer high torque capabilities at low speeds, provide high efficiencies over the entire speed range and reduce noise emissions \cite{rasmussen_1997}. For those reasons, they are emerging as the most prevalent VSD refrigeration technology \cite{Samsung_2015}.

Here, a three-phase BLDC operating in a two-phase conduction mode is considered \cite{kim_chap2_2017,krishnan_permanent_2017}. The basic structure is similar to the one of induction motors. Namely, the three-phase armature windings are located on the stator and the excitation system is placed on the rotor, with the magnetic coupling realised via permanent magnets on the rotor. Despite the similarities in their layout, the operating principles differ. In particular, the BLDC's armature windings are energised with rectangular DC voltage pulses in accordance with the instantaneous rotor position, thereby imposing torque and enabling energy transfer. 

A block diagram of the BLDC model in frequency domain is displayed in Fig.~\ref{fig:BLDC}, with the motor dynamics described by
\begin{subequations}
\label{eq:BLDC1_2}
\vspace{-5pt}
\begin{align}
    \dot{i}_m &= \frac{\omega_{b}}{\ell_a} \, \bigl(v_{m,2}-r_a i_m - \overbrace{k_e \, \omega_m}^{\text{EMF}} \,\bigr), \label{eq:BLDC1}\\ 
    \dot{\omega}_m &= \frac{1}{2H_m}\, \bigl(\,\underbrace{k_t i_m}_{t_e}-t_c - b\, \omega_m\bigr). \label{eq:BLDC2}
\end{align}
\vspace{-2pt}
\end{subequations}
The armature current $i_m$ depends on the armature coil parameters $(r_a, \ell_a)$, the base frequency $\omega_{b}$, the modulated input DC voltage $v_{m,2}$, and the back EMF, determined by the rotational speed $\omega_m$ and the EMF constant $k_e$. Furthermore, a swing equation relates the mismatch in electromagnetic $t_e$ and load $t_c$ torque to the change in rotational speed, proportional to the rotor inertia constant $H_m$. The electromagnetic torque is considered to linearly depend on the armature current, described by the torque coefficient $k_t$. In an ideal motor, the numerical value of this coefficient equals the EMF constant in SI units~\cite{kim_chap2_2017}. Finally, the friction is also included in the model through a viscous friction coefficient $b$. 

It should be noted that a BLDC is usually driven by a controlled converter that regulates the required switches depending on the instantaneous rotor position. In a real-world implementation, the rotor position is monitored by the use of Hall-sensors or other sensorless schemes for rotor position estimation~\cite{kim_brushless_2017}. The sensorless control schemes particularly prevail in refrigeration applications to limit the number of connections in the hermetic compressor housing~\cite{xia_permanent_2012}. However, for the purposes of this study we will assume that the rotor angle measurement and electrification of the corresponding phases operates with sufficient accuracy and reliability.

\subsection{Variable-Speed Drive}
A VSD supplies the required rectangular DC voltage to the BLDC and controls the shaft speed. In this paper, we implement a configuration with two full-bridge converters (rectifier and inverter) with common constant DC-link voltage. 

\subsubsection{Inverter}
\begin{figure}
    \centering
    \includegraphics{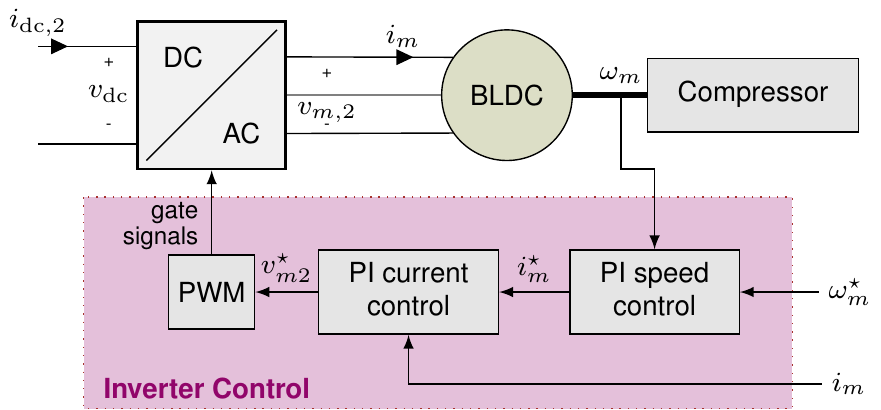}
    \caption{Inverter control scheme.}
    \label{fig:Inverter}
\end{figure}
The inverter adjusts the rotational speed of the BLDC by regulating its input voltage ($v_{m}$) through Pulse-Width Modulation (PWM) and controlling the motor current $(i_m)$. Based on \cite{kim_brushless_2017}, a sequence of two PI-control loops is employed, as illustrated in Fig.~\ref{fig:Inverter}.

First, the speed PI-controller determines the required single-phase BLDC peak current reference $i_m^\star$ according to the mismatch between the rotational speed $\omega_m$, its reference $\omega_m^\star$ and the instantaneous motor current $i_m$. The control law in time domain yields
\begin{subequations}
\begin{align}
    i_{m}^\star &= i_{m} + k_{ps}\, \bigl( \omega_m - \omega_m^\star \bigr) + k_{is}\, \mu_{\omega_m}, \\
    \dot{\mu}_{\omega_m} &= \omega_m-\omega_m^\star, 
\end{align}
\end{subequations}
where $k_{ps}$ and $k_{is}$ are the proportional and integral gain of the speed controller, respectively. The second PI-controller regulates the modulation voltage reference $v_{m,2}^\star$ for the motor current $i_m$ to match the internal reference signal, described by the control law of the form:
\begin{subequations}
\begin{align}
    v_{m,2}^\star &= v_\mathrm{dc} + k_{pc,2}\, \bigl( i_{m} - i_{m}^\star \bigr) + k_{ic,2}\, \mu_{i_m}, \\
    \dot{\mu}_{i_m} &= i_{m}-i_m^\star,
\end{align}
\end{subequations}
with $k_{pc,2}$ and $k_{ic,2}$ denoting the appropriate proportional and integral control gains.

Finally, the PWM adjusts the modulation index such that the energized phases of the BLDC receive the desired voltage. Here, the operation of the PWM is assumed ideal, and thus it is not further modelled, ergo $v_{m,2}^\star = v_{m,2}$. Finally, the DC input current of the inverter $i_{\mathrm{dc},2}$ is determined by exploiting the power balance across the converter and assuming a lossless operation, i.e., 
\begin{align}
    i_{\mathrm{dc},2} &= \frac{v_{m,2} i_m}{v_\mathrm{dc}}. \label{eq:powerBalance}
\end{align}

\subsubsection{Rectifier}
\begin{figure}[t]
    \centering
    \includegraphics{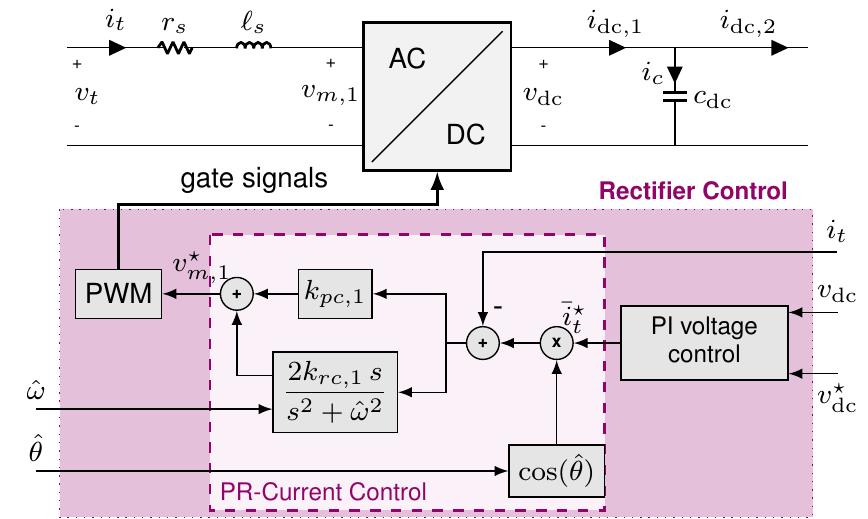}
    \caption{Conventional PR-control for active single-phase rectifiers.}
    \label{fig:Rectifier}
\end{figure}

The rectifier maintains the common DC-link voltage constant, regulates the terminal current $i_t$ and keeps a unity power factor at the PCC. Two sequential controllers are usually employed, namely a PI-controller regulating the DC voltage followed by a current controller. The current control can also be implemented as a PI-control on the AC-terminal current error. However, it is mostly used in low-cost applications due to an inevitable steady-state error. In contrast, a Proportional-Resonant (PR) control is capable of successfully tracking the sinusoidal current reference~\cite{hong-seok_song_advanced_2003}. Therefore, the latter approach is considered in this paper, as shown in Fig.~\ref{fig:Rectifier}. Note that for the purposes of small-signal analysis, the mathematical representation of the given model requires modification due to the sinusoidal shape of the current control reference. 

\begin{figure}
    \centering
    \includegraphics{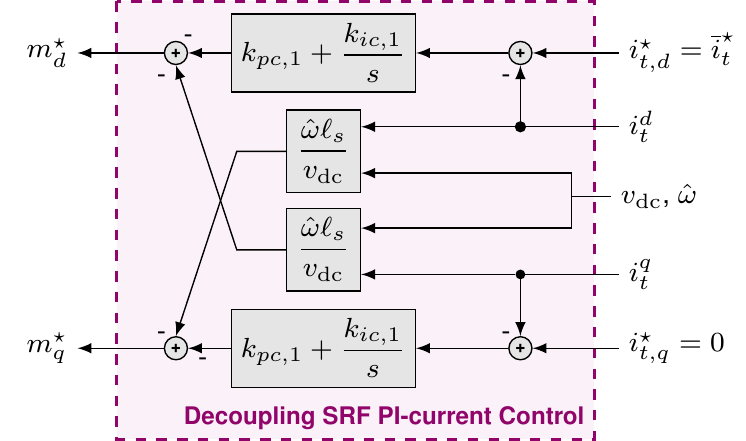}
    \caption{Decoupling SRF current control scheme.}
    \label{fig:SRF_Rectifier}
\end{figure}

The equivalence between a PR-current control and a decoupling PI-current control in an SRF has been shown in~\cite{bacha_power_2014, hackl_equivalence_nodate}. In other words, controlling the $dq$-components of the park-transformed terminal current in a decoupled fashion is equivalent to employing a PR-control on the respective phases in a stationary reference frame. Consequently, the PR-current control is replaced by decoupling current control (see Fig.~\ref{fig:SRF_Rectifier}) and the entire grid-side AC system is modeled in the appropriate SRF. Nonetheless, the conclusions drawn in this work still hold for systems using resonant controllers. In the following, the mathematical model of the rectifier control is presented.


The PI-control processes the mismatch of the DC-link voltage $v_\mathrm{dc}$ and the external voltage reference $v_\mathrm{dc}^\star$. Thereby, it determines the magnitude of the terminal current $\overline{i}_t^\star$, equivalent to the $d$-component of the terminal current reference ${{i}_t^{d}}^\star$:
\begin{subequations}
\begin{align}
    {i_t^d}^\star &= k_{pv}\, \bigl( v_\mathrm{dc}^\star - v_\mathrm{dc} \bigr) + k_{iv}  \mu_v ,\\
    \dot{\mu}_v &= v_\mathrm{dc}^\star-v_\mathrm{dc},
\end{align}
\end{subequations}
with $k_{pv}$ and $k_{iv}$ indicating the proportional and integral control gains, respectively. To preserve a unity power factor at the terminal, the current reference must be in phase with the mains voltage. Hence, it is multiplied by the estimated phase shift $\cos(\hat{\theta})$ within the PR-control design. Correspondingly, the $q$-axis component of the terminal current reference $i_{t,q}^\star$ must be zero. 

Subsequently, the terminal current reference is passed to the decoupling terminal current PI-control, described by
\begin{subequations}
\begin{align}
    {m}_{dq}^\star &= -k_{pc,1} ({i}_{t,dq}^\star - i_{t}^{dq}) - k_{ic,1} {\mu}_c^{dq} - j \frac{\ell_s \hat{\omega}}{v_\mathrm{dc}^\star} i_{t}^{dq},\\
    \dot{{\mu}}_{c}^{dq} &= {i}_{t,dq}^\star - i_t^{dq},
\end{align}
\end{subequations}
where $k_{pc,1}$ and $k_{ic,1}$ are the respective proportional and integral control gains, ${m}_{dq}^\star$ is the modulation reference for the PWM, and $\hat{\omega}$ is the Phase-Locked Loop (PLL) estimate of the mains frequency. Note that for the equivalence of the two discussed controllers, the resonant control gains must be equal to the SRF-PI current control gains, namely $k_{rc,1}=k_{ic,1}$. Similar to the inverter control design, the PWM of the rectifier is assumed to have ideal operation and therefore ${m}_{dq}^\star = {m}^{dq}$ holds.

\subsubsection{Electrical System Dynamics}
The VSD is connected to the grid via an impedance $(r_s,\ell_s)$. Hence, the average terminal current dynamics in time domain and SRF are of the form:
\begin{align}
     \dot{{i}}^{dq} &= -{j} \omega_g \omega_b {i}^{dq} + \frac{\omega_b}{\ell_s}\left({v}^{dq} - {m}^{dq} v_\mathrm{dc}-r {i}^{dq} \right),
\end{align}
as shown in \cite{yang_modeling_2018}.
Moreover, the common DC-side dynamics across the DC-link capacitor $c_\mathrm{dc}$ can be derived as
\begin{align}
    \dot{v}_\mathrm{dc} &=\frac{3}{2} \frac{\omega_b}{c} \left( m^d i^d + m^q i^q\right) - \frac{\omega_b}{c} i_{\mathrm{dc},2}.
\end{align}

\subsection{Single-Phase PLL}
\begin{figure}[t]
    \centering
    \includegraphics{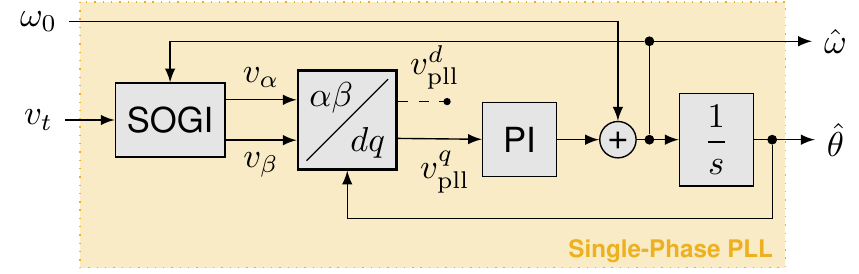}
    \caption{Single-phase SOGI PLL.}
    \label{fig:Sogi_pll}
\end{figure}
\begin{figure}[t]
    \centering
    \includegraphics{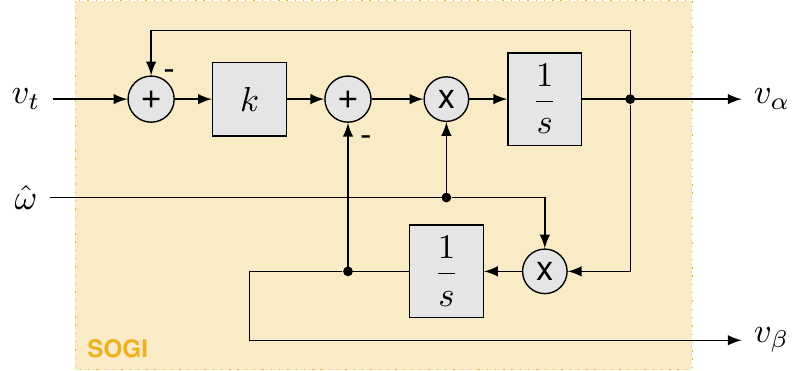}
    \caption{Block diagram of the SOGI module.}
    \label{fig:SOGI}
\end{figure}

For the grid synchronization unit, the standard Second-Order General Integrator PLL (SOGI-PLL) proposed in \cite{ciobotaru_2006} is considered, illustrated in Fig.~\ref{fig:Sogi_pll}. The SOGI-PLL observes the terminal voltage and tracks the mains frequency by internally generating an orthogonal system $(\alpha\beta)$ from voltage measurements, transforming it to an internal $(dq)$ SRF, and passing the phase angle difference through a PI-control \cite{xiao_2017, golestan_2013}.

A mathematical formulation of the orthogonal system generator in frequency domain is presented in Fig.~\ref{fig:SOGI}. The orthogonal terminal voltages in stationary reference frame $(v_\alpha,v_\beta)$ are established by second-order integration and filtering, where the estimated mains frequency $\hat{\omega}$ is used as the resonant frequency and the gain factor $k$ determines the level of filtering around that frequency. With decreasing gain factor, the bandpass narrows while at the same time increasing the robustness against disrupted input signals \cite{ciobotaru_2006}.

The authors in \cite{xiao_2017, golestan_2013} derive a small-signal representation of the SOGI-operation that is valid under certain assumptions: (i) there is a small phase angle difference between the grid angle $\theta$ and the estimated angle $\hat{\theta}$; and (ii) the estimated frequency $\hat{\omega}$ is close to the real frequency $\omega_g$. If these conditions are met, in frequency domain one obtains 
\begin{align}
    V_\mathrm{pll}^q(s) &= \frac{1}{\frac{2}{k{\omega_g}} s+1} \left[ \theta(s) -\hat{\theta}(s) \right].
\end{align}
A Laplace transform yields the desired mathematical formulation of the SOGI operation in time domain:
\begin{align}
    \dot{v}_\mathrm{pll}^q &= \frac{k {\omega_g}}{2} \left( \theta - \hat{\theta} - v_\mathrm{pll}^q \right).
\end{align}
Subsequently, a PLL estimates the frequency $\hat{\omega}$ and the voltage phase angle $\hat{\theta}$ at the terminal, described by
\begin{subequations}
\begin{align}
    \hat{\omega} &= k_{p,pll}\,v_\mathrm{pll}^q + k_{i,pll} \mu_\mathrm{pll}  + \omega_0, \label{eq:PLLPI}\\
    \dot{\mu}_\mathrm{pll} &= v_\mathrm{pll}^q, \label{eq:PLL} \\
    \dot{\hat{\theta}} &= \hat{\omega} \omega_b.
\end{align}
\end{subequations}
 Here, $k_{p,pll}$ and $k_{i,pll}$ represent the proportional and integral PI-control gains respectively, whereas $\omega_0$ is the nominal grid frequency.

\subsection{Speed Reference Computation}
Considering that the focus of this work is VSDR contribution to FFC, the speed reference $\omega_m^\star$ is computed by achieving a trade-off between two control objectives and timescales. In the long term, the temperature in the refrigeration compartment must not deviate significantly from its external reference in order not to impair the user comfort. In the short term, the FFR provision requires that the rotational speed reacts to frequency disturbances and deviates from its original reference. We combine the two requirements as follows:
\begin{align}
    \omega_m^\star &= \omega_{m,T}^\star + \Delta\omega_m^\star,
\end{align}
where $\Delta\omega_m^\star$ and $\omega_{w,T}^\star$ are the required contributions for providing frequency support and controlling the cool chamber temperature, respectively.

\begin{figure}
    \centering
    \includegraphics{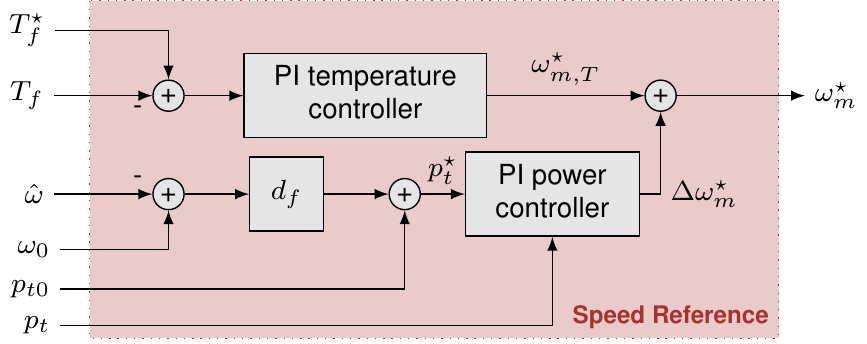}
    \caption{Computation of the speed reference.}
    \label{fig:speedReference}
\end{figure}

\subsubsection{Temperature Control}
In conventional VSDR, the speed reference is mostly driven by the temperature deviation in the refrigeration compartment \cite{li_feedforward_2009}. In this paper, the slow variation of the rotational speed reference $\omega_{m,T}^\star$ of the BLDC is determined by the deviation of the refrigeration compartment temperature $T_f$ from its external reference $T_f^\star$. The corresponding contribution of the PI-control can be described by
\begin{align}
    \omega_{m,T}^\star &= k_{pT}\, (T_f^\star - T_f) + k_{iT} \mu_T , \\[5pt]
    \dot{\mu}_T &=  T_f^\star-T_f ,
\end{align}
where $k_{pT}$ and $k_{iT}$ are the proportional and integral gains.

\subsubsection{Active Power Control}
To operate in grid-supporting mode, the compressor speed must be adjusted according to measured frequency deviations. Based on the converter classification and control guidelines in \cite{rocabert_2012}, the frequency droop is applied to calculate the new power reference $p_t^\star$ and a PI-controller determines the required compressor speed deviation $\Delta \omega_m^\star$ based on the mismatch between the power consumption $p_t$ and the reference. The mathematical formulation is given by
\begin{align}
    p_t &= \frac{1}{2} \left(v_t^d \, i_t^d+v_t^q \, i_t^q \right) \\
    p_t^\star &=  p_{t0} + d_f\, (\omega_0 -\hat{\omega} ) \label{eq:frequencyDroop}, \\[5pt]
    \Delta\omega_m^\star &= k_{pp}\, (p_t^\star -p_t) + k_{ip} \mu_{pt} ,  \label{eq:pCtrl}  \\[5pt]
    \dot{\mu}_{pt} &=p_t^\star-p_t.
\end{align}

Here, $d_f$ refers to the frequency droop gain, $\omega_0$ to the nominal grid frequency, $p_{t0}$ to the power consumption prior to the disturbance, while $k_{pp}$ and $k_{ip}$ are the proportional and integral PI-control gains, respectively. 

\subsection{Grid Equivalent} \label{sec:GridEquivalent}
\begin{figure}
    \centering
    \includegraphics{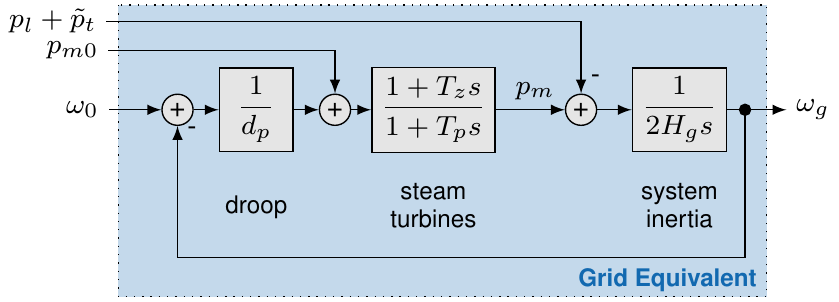}
    \caption{Block diagram of the implemented grid equivalent.}
    \label{fig:gridEquivalent}
    \vspace{-5pt}
\end{figure}
To analyse the effect of VSDR control on grid frequency, the grid equivalent is implemented according to \cite{weckesser_2017,misyris_2019} and detailed below. Besides providing the voltage input to the PLL, the major purpose of grid equivalent is to capture the electromechanical dynamics of the centre-of-inertia frequency after a disturbance. The model depicted in Fig.~\ref{fig:gridEquivalent} reflects the inertial response and primary frequency control provided predominantly by conventional steam turbines \cite{weckesser_2017}. The mathematical representation of the given model is of the form:
\begin{subequations}
\begin{align}
   \dot{p}_m = & - \frac{1}{T_p} (p_m -p_{m0}) - \frac{1}{d_p T_p} \overbrace{\left(\omega_g-\omega_0\right)}^{\Delta \omega_g} \nonumber\\
    & - \frac{T_z}{2H_g d_p T_p} (p_m -p_l-\tilde{p}_t), \label{eq:pm}\\  
    \Delta\dot{\omega}_g = & \,\frac{1}{2H_g} (p_m-\tilde{p}_t-p_l), \label{eq:dw}\\
    \dot{\theta}_g = & \,\omega_g \omega_b. \label{eq:thetag}
\end{align}
\end{subequations}
While $p_{m0}$ and $p_{m}$ denote the initial and instantaneous mechanical power output of the steam turbine, $p_l$ and $\tilde{p}_t$ represent the power demand and aggregated load of $n$ VSDR units, respectively. The time constants of the steam turbine $(T_p, T_s)$, the droop constant $d_p$ and the system inertia constant $H_g$ characterise the inertial and primary frequency response~\cite{weckesser_2017}.

In addition to the electromechanical dynamics, the grid equivalent also provides a stiff terminal voltage
\begin{subequations}
\label{eq:vt_dq}
\begin{align}
    {v}_t^{d} &= x_g {i}_t^{q} + v_g   \cos(\theta_g-\hat{\theta}), \\
    {v}_t^{q} &= - x_g {i}_t^{d} + v_g   \sin(\theta_g-\hat{\theta}),
\end{align}
\end{subequations}
where $v_g$ is the constant grid voltage magnitude, $\theta_g$ is the phase of the voltage according to (\ref{eq:thetag}), and $x_g$ is the internal impedance of the grid. 

Note that the VSDR aggregation power output represents a summation of individual outputs of $n$ active units, i.e., $\tilde{p}_t = n p_t$. Such aggregation method was established in \cite{Mahdavi20} for inverter air conditioners and can be applied to VSDR due to similarities in their operation. Moreover, \cite{Mahdavi20} verifies that a sample average unit model provides an accurate representation of the entire aggregation. This is true even when the physical system parameters (e.g., thermal parameters, internal setpoints and control loop parameters) are heterogeneously distributed. Thus, the parameterisation provided in Table~\ref{tab:parameters} reflects the average parameters of an aggregation of devices.
\begin{table}[!t]
\processtable{Detailed Model Parameters.\label{tab:parameters}}
{\begin{tabular*}{20pc}{@{\extracolsep{\fill}}l l p{0pt} l l@{}}
\toprule
\multicolumn{2}{l}{\scriptsize \textbf{BLDC Parameters}}  & 	&        \multicolumn{2}{l}{\scriptsize\textbf{Calculation of Speed Reference}}  \\  \cmidrule{1-2} \cmidrule{4-5}
      $r_a$& \SI{0.0081}{\pu} & & $k_{pT}$ & \SI{-0.159}{\per\kelvin}\\
         $l_a$ & \SI{0.015}{\pu}& &  $ k_{iT}$ & \num{-3.18e-5} $\mathrm{(Ks)}^{-1}$\\
         $H_m$ &\SI{0.2023}{\second} &	  &  $d_f$ &\SI{20}{\second\per\radian} \\
         $b$ & \SI{0.0987}{\pu} &	  & $k_{pp}$    & \SI{4.5}{\pu}\\
         $k_t = k_e$ & \SI{0.7398}{\pu}&	& $k_{ip}$ & \SI{90}{\per\second}\\
         rated speed & \SI{3000}{rpm} & & & \\ \cmidrule{1-2} \cmidrule{4-5}
  \multicolumn{2}{l}{\scriptsize\textbf{Electric Components}}  &	    &     \multicolumn{2}{l}{\scriptsize\textbf{SOGI-PLL}}    \\ \cmidrule{1-2} \cmidrule{4-5}
         $c$ & \SI{11.43}{\pu} & &    $k_{p,pll}$ & \SI{0.4}{\radian\per\second} \\  
         $r_s$ & \SI{0.012}{\pu} &	  &    $k_{i,pll}$   & \SI{4.69}{\radian\per\second}\\
         $l_s$ & \SI{0.038}{\pu} & &  $k$ & \num{1.63} \\ \cmidrule{1-2} \cmidrule{4-5}
         \multicolumn{2}{l}{\scriptsize\textbf{Refrigerator Model}}  &	   &   \multicolumn{2}{l}{\scriptsize\textbf{Grid Equivalent}}  \\ \cmidrule{1-2} \cmidrule{4-5} 
         $a_2$ & \num{-0.295}& &  $P_g$ & \SI{200}{\mega\watt} \\ %
          $a_1$ & \num{1.583} & &  $\omega_b$ & \SI{314.16}{\radian\per\second}\\  %
          $a_0$ & \num{-0.075}& &     $H_g$ & \SI{0.5}{\second}\\  
         $b_1$  &  \num{-1.64e-5}  & &  $T_z,T_p$ & \SI{2.1}{\second}, \SI{7}{\second}\\ 
         $b_3$ &   \num{0.558} &	   &  $d_p$ & \SI{0.02}{\pu}\\
         $b_2, b_4$ & \num{5.909}, \num{0.086}  & &   $v_g$ &  \SI{1.41}{\pu}\\ 
         $T_a, T_f^\star$& \SI{32}{\degreeCelsius}, \SI{3}{\degreeCelsius}  &	  & $p_{l0}$ & \SI{1}{\pu} \\
	    $\tau_q, \tau_p$ & \SI{100}{\second}, \SI{1}{\second} &  & $x_g$ & \SI{0.15}{\pu}\\  \cmidrule{4-5}
	    $r_{th}$ &  \SI{55}{\kelvin}& &  \multicolumn{2}{l}{\scriptsize\textbf{VSD Control}}  \\ \cmidrule{4-5}
        $ c_{th}$ & \SI{454.6}{\kelvin} &    &$k_{pc2}, k_{ic2}$ & \num{0.019}, \SI{3.226}{\per\second}\\
        $P_b$ &\SI{100}{\watt}  & 	& $k_{pv}, k_{iv}$ &\num{4.973}, \SI{239.7}{\per\second} \\
        $ $    & $ $ &	    &           $k_{pc1}, k_{ic1}$ & \num{20.59}, \SI{1672}{\per\second}\\
         & & &  $k_{ps}, k_{is}$ & \num{43.76}, \SI{700}{\per\second} \\ \bottomrule
\end{tabular*}}{}
\end{table}
\subsection{Full-Order State-Space Model}
The resulting DAE model can be written in the following form: 
\begin{align}
    \begin{split}
        \dot{x} &= f(x,y,u), \\
        0 &= g(x,y,u), \label{eq:stateSpace} 
    \end{split}
\end{align}
where the first-order differential equations $f(x,y,z)$ and the algebraic equations $g(x,y,u)$ link the vectors of state variables $x$, algebraic variables $y$ and input variables $u$. 

By linearizing the model around its stationary operation point, the state-space representation and small-signal model is obtained as
\begin{subequations}
\label{eq:full_sys}
\begin{align}
    \dot{{x}} &= {A} {x} + {B} {u}, 
\end{align}
with ${A}$ being the state-space matrix and ${B}$ the input-state matrix. The complete model includes expressions \eqref{eq:qth_pc}-\eqref{eq:vt_dq} and comprises 21 state variables and 9 control inputs:
\begin{align}
\begin{split}
     {x} =  \bigl[&T_f, \omega_m, i_m, t_c, q_{th}, i^d, i^q,  v_\mathrm{dc}, \hat{\theta}, \theta_g, v_\mathrm{pll}^q, \ldots\\
     &p_m, \Delta\omega_g, \mu_c^d, \mu_c^q, \mu_T, \mu_v, \mu_{\omega_m}, \mu_{im}, \mu_\mathrm{pll}, \mu_{pt}\bigr]^\mathsf{T}, 
\end{split} \\[5pt] 
    {u} = \bigl[&p_l, T_f^\star, v_\mathrm{dc}^\star,{i^q}^\star, T_a, p_{t0}, p_{m0}, w_0, v_g \bigr]^\mathsf{T}. \label{eq:u}
\end{align}
\end{subequations}

The most relevant model parameters are listed in Table~\ref{tab:parameters}. It should be noted that the power and speed control parameters were manually tuned, whereas the BLDC PI-control loops were tuned according to the technique proposed in \cite{kim_chap2_2017}. Moreover, the rectifier parameters and control gains were taken from \cite{yang_modeling_2018} and scaled accordingly, while the PLL parameterisation was obtained from \cite{markovic_stability_2018}.

\section{Simplified VSDR Model}\label{sec3}
In power systems literature, the VSD-based technologies (e.g., refrigeration, air conditioning and heat pumps) are usually represented by simplified dynamic models to assess the overall response of an aggregation of devices with the same technology. In particular, \cite{Malekpour20} and \cite{Ibrahim20} include the induction machine inertia, torque control and DC-link dynamics and derive a third-order model for a VSD heat pump system. The authors in \cite{kim_modeling_2015} model the electrical side of VSD heat pumps by employing first-order motor dynamics and a first-order transfer function for the VSD, whereas \cite{hui_equivalent_2019} derives a third-order model for an inverter air conditioning system. However, none of the aforementioned studies analyses the accuracy of the simplified models nor justifies the level of modelling detail.

Considering the high computational burden pertaining to modelling a significant number of different device aggregations in high detail presented in Section~\ref{Sec2}, in this section we focus on deriving an appropriate low-order transfer function. Therefore, we consider a detailed DAE model comprising VSD, motor, AC side and refrigeration dynamics and subsequently subsume it within a single transfer function. As a result, the model order of a single device or ensemble of units is significantly reduced, thus enabling large-scale simulations with acceptable computation times. Simultaneously, the derived simplified models preserve the physical system states that are of high importance to system operators, namely the rotational speed of the BLDC that can be used for reserve estimation and the power output of a single device.

\begin{figure}
    \centering
    \includegraphics{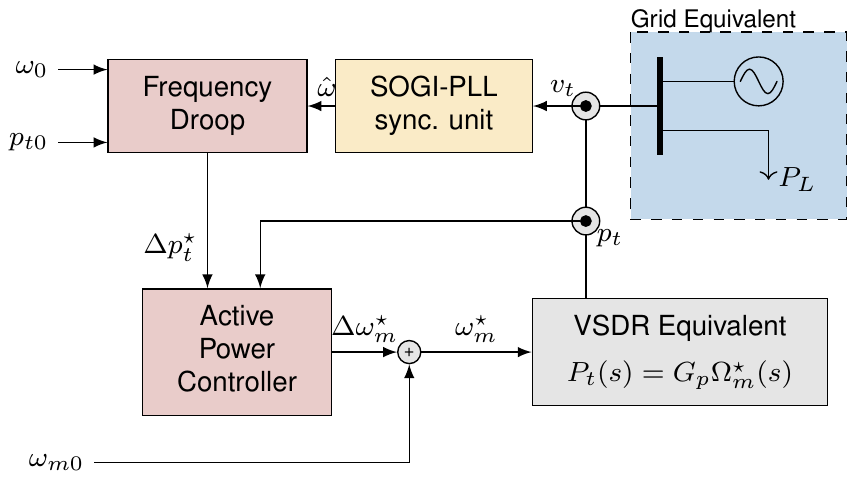}
    \caption{Overview of the simplified model.}
    \label{fig:SimplifiedModel}
\end{figure}

An overview of the simplified control structure is provided in Fig.~\ref{fig:SimplifiedModel}. While all electric and thermal dynamics are subsumed in the transfer function $P_t(s)$, the SOGI-PLL, frequency droop, active power control and grid equivalent remain unchanged. In the remainder of this section, the simplified model formulation is presented and the respective transfer functions are tuned using the full-order model time-domain simulation results.

\subsection{Simplified VSDR Model Formulation}

The simplified model represents the VSDR using transfer functions of varying order (up to third) that map the BLDC's speed reference to the power consumption at the device terminal, as follows:
\begin{align}
    p_t(s) &= \frac{n_2s^2 + n_1 s + n_0}{s^3 + d_2 s^2+d_1 s + d_0} \omega_m^\star,
\end{align}
which transformed into state-space form yields
\begin{align}
        \dot{{v}} &= \left[\begin{matrix} -d_2 & -d_1 & -d_0  \\ 1 & 0 & 0 \\ 0 & 1 & 0 \end{matrix}\right] {v} + \left[\begin{matrix} 1 \\0 \\ 0 \end{matrix}\right] \omega_m^\star, \\
        p_t &= \left[\begin{matrix} n_2 & n_1 &  n_0 \end{matrix} \right] {v},
\end{align}
where $d_2, d_1, d_0$ are the denominator constants, $n_2, n_1, n_0$ are the nominator parameters, and ${v}$ is the state vector of the respective transfer function. 

The simplified model reduces the order of the system in \eqref{eq:full_sys} from 21 to 8-10, depending on the order of the transfer function. The reduced-order state-space formulation is thus of the form:
\begin{align}
\begin{split}
     x &= \bigl[v_{1,2,3}, \hat{\theta}, \theta_g, v_\mathrm{pll}^q, p_m, \Delta \omega_g, \mu_\mathrm{pll}, \mu_{pt} \bigr]^\mathsf{T}, \\
     u &= \bigl[p_l, p_{t0}, p_{m0}, \omega_0\bigr]^\mathsf{T}.
\end{split}
\end{align}

\subsection{Transfer Function Fitting}
\begin{table}[!t]
\processtable{Transfer function fitting results and simplified model parameters. The relative fit of the transfer function is given below each model.\label{tab:TFparameters}}
{\begin{tabular*}{0.95\linewidth}{@{\extracolsep{\fill}} c  S[table-format=3.1(1)e2] S[table-format=3.1(1)e2] S[table-format=3.1(1)e2] @{}}
  \toprule 
   &{\textbf{\scriptsize P3Z2}}   &  {\scriptsize\textbf{P3Z1}} & {\scriptsize\textbf{P3Z0}} \\  
   & {\tiny \SI{77}{\percent} fit}   &    {\tiny \SI{77}{\percent} fit}  &  {\tiny \SI{22}{\percent} fit}  \\ \cmidrule{2-2} \cmidrule{3-3} \cmidrule{4-4}
  $n_2$ & -454.27  & 0  &       0         \\ 
  $n_1$ & 3.879e6 &  3.456e6 & 0         \\ 
  $n_0$ & 7.955e6 &  7.084e6 &  1.318e11 \\ 
  $d_2$ & 4.332e3     & 3.878e3   & 3.966e5    \\ 
  $d_1$ & 1.994e5  &   1.778e5    & 8.833e7  \\ 
  $d_0$ & 1.065e7 & 9.480e6   & 1.745e11 \\ \cmidrule{2-2} \cmidrule{3-3} \cmidrule{4-4}
&    {\scriptsize \textbf{P2Z1}}   	&      {\scriptsize\textbf{P2Z0}}  & {\scriptsize\textbf{P1Z0}} \\  
 &    {\tiny \SI{77}{\percent} fit}   & {\tiny \SI{23}{\percent} fit}  &  {\tiny \SI{22}{\percent} fit} \\ \cmidrule{2-2} \cmidrule{3-3} \cmidrule{4-4}
  $n_2$ & 0          & 0       & 0 \\
  $n_1$ & 890.01     & 0       & 0\\ 
  $n_0$ & 1.83e3    & 3.519e3  & 731.36\\ 
  $d_2$ & 1          & 1       & 0 \\
  $d_1$ & 45.14      & 6.169   & 1\\ 
  $d_0$ & 2.43e3 & 4.651e3     & 964.8\\ \bottomrule
\end{tabular*}}{}
\end{table}

The transfer function parameters were obtained by applying 10 step changes in reference speed to the full DAE model, covering the entire range of BLDC operation from \SIrange{1000}{4000}{rpm}. For this purpose, the active power controller was disabled and a direct step change in the speed reference was applied to the BLDC controller. The simulated power response was then fitted to several transfer functions using the System Identification Toolbox in MATLAB, with the obtained parameters given in Table~\ref{tab:TFparameters}. For simplicity, different transfer functions are denoted by $PiZj$, where $i$ refers to the number of poles and $j$ to the number of zeros. 
\begin{figure}
    \centering
    \includegraphics{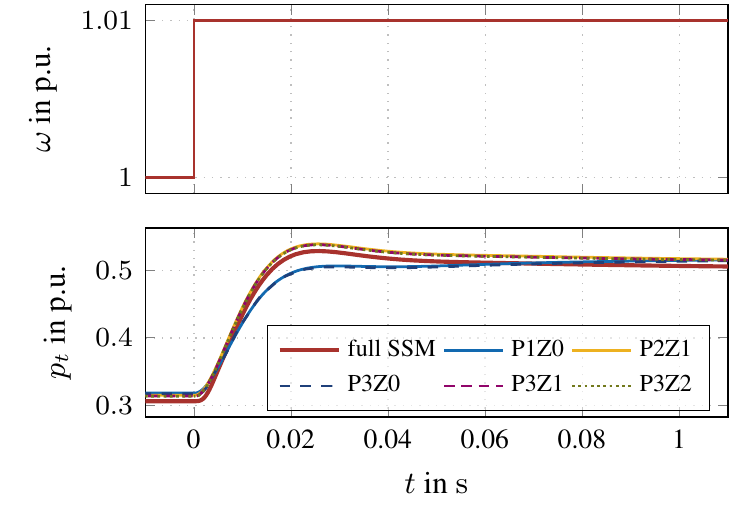}
    \caption{Closed-loop system response of the full-order SSM and reduced-order models when subjected to a step change in mains frequency and the droop controller enabled. The model $P2Z0$ is unstable and therefore not included.}
    \label{fig:Tuning2}
\end{figure}
For validation of the transfer function fitting, two test setups were considered and their time-domain performance was compared. First, the models under investigation were subjected to the aforementioned sequence of ten step changes with the droop control disabled. Subsequently, the VSDR droop control was enabled and the time-domain response to a sample disturbance in the mains frequency was observed. Since the analysis of both test cases arrives at similar conclusions, the latter test case will be showcased and analysed here as it is more relevant to the problem at hand.

The time-domain performance of the detailed SSM and its reduced equivalents is investigated for a step change in grid frequency of $\Delta\omega_g =\SI{0.01}{\pu}$ at $t=\SI{0}{\second}$ and depicted in Fig.\ref{fig:Tuning2}. The model $P2Z0$ is not included since it exhibits instability for the given droop parameters. Overall, the models containing at least one additional zero (i.e., $P2Z1$, $P3Z1$ and $P3Z2$) outperform the other stable models in terms of transient behaviour. While experiencing a small but constant offset in estimation of the power at the terminal, these models capture the overall trajectory of the transient power response better. In contrast, the two stable models without additional zeros (i.e., $P1Z0$ and $P3Z0$) seem to reach the post-fault steady-state faster, but are not able to accurately capture the overshoot in terminal power. Similar behaviour was observed for other step changes in mains frequency and rotational frequency reference. The following sections further examine the closed-loop system response as well as the interactions with the grid.

\section{Performance and Model Validation} \label{sec4}
\begin{figure}[b]
    \centering
    \includegraphics{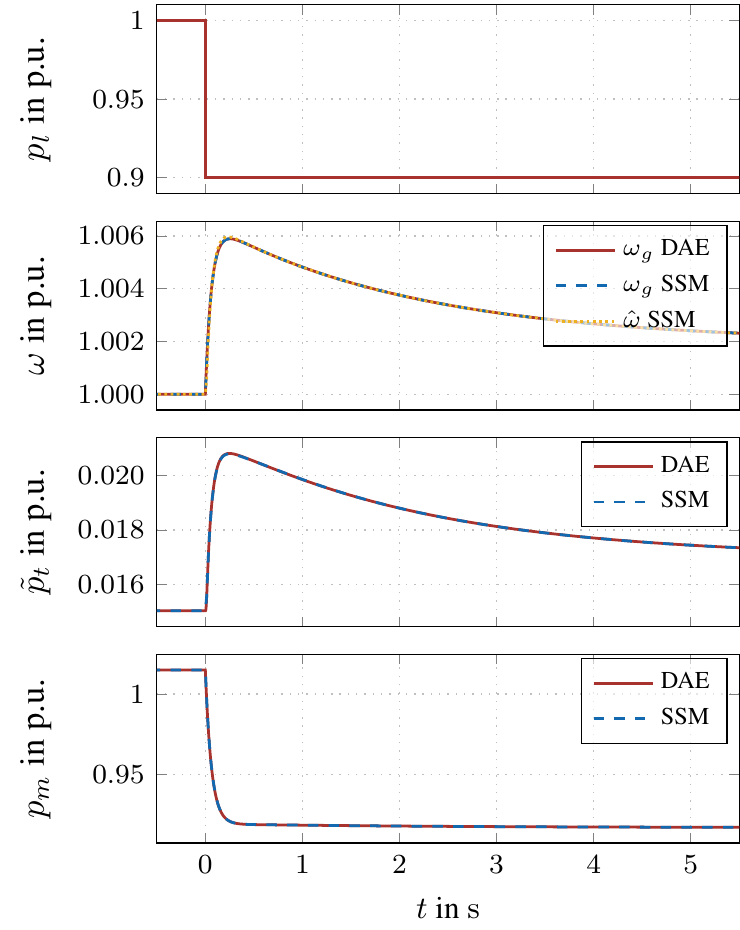}
    \caption{Aggregated power response and frequency estimation of the VSDR unit at the grid level when subjected to a step change in load power.}
    \label{fig:GridResponse}
\end{figure}
\begin{figure}[b!]
    \centering
    \includegraphics{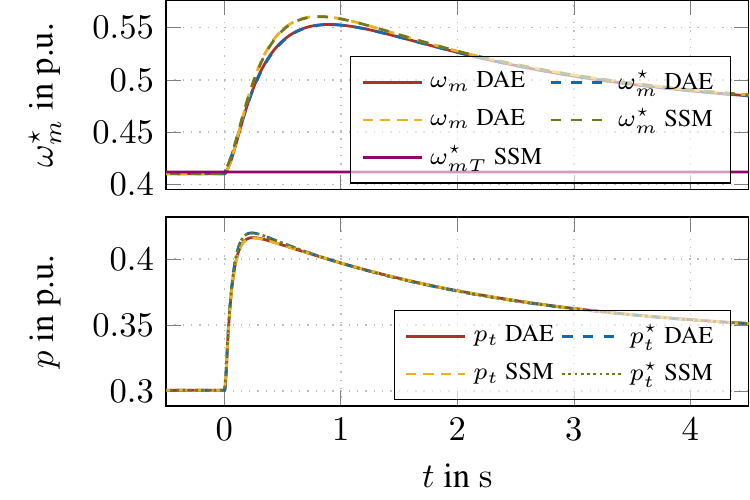}
    \caption{Terminal power, computation of the speed reference and rotational speed of a single VSDR unit after a step change in load power.}
    \label{fig:SingleDevice}
    \vspace{-10pt}
\end{figure}
This section examines the closed-loop system response and grid interactions of both the full-order and the reduced-order small signal models. For validation of the SSM, the time-domain performance of individual system components is investigated and compared to the benchmark DAE model solution. Subsequently the accuracy of all reduced-order models is quantified.

The systems are subjected to a step decrease in power system background load power of $0.1\,\mathrm{p.u.}$ at time instance of $t=0\,\mathrm{s}$. An aggregation of $100\,000$ VSDR devices is assumed that employs a stabilising frequency droop of $d_f=20\mathrm{\,s}$, corresponding to a power droop of $\SI{5}{\percent}$. Compared to the base power, this aggregation equates to a refrigeration capacity of \SI{5}{\percent} and is comparable to the actual share of refrigerators in electricity consumption of Switzerland \cite{Kemmler18}. A starting reference speed for BLDC is set to $\omega_{m,T}^\star = 0.41\,\mathrm{p.u.}$, which is the speed at which the refrigeration compartment temperature is kept constant for the given set of refrigeration parameters. As explained in Section\ref{sec:GridEquivalent}, the initial steady state and parameterisation of the single VSDR considered here reflects the average parameters and behaviour of the ensemble.

\subsection{Time-Domain Performance of Full-Order Model}
\begin{figure}[b]
    \centering
    \includegraphics{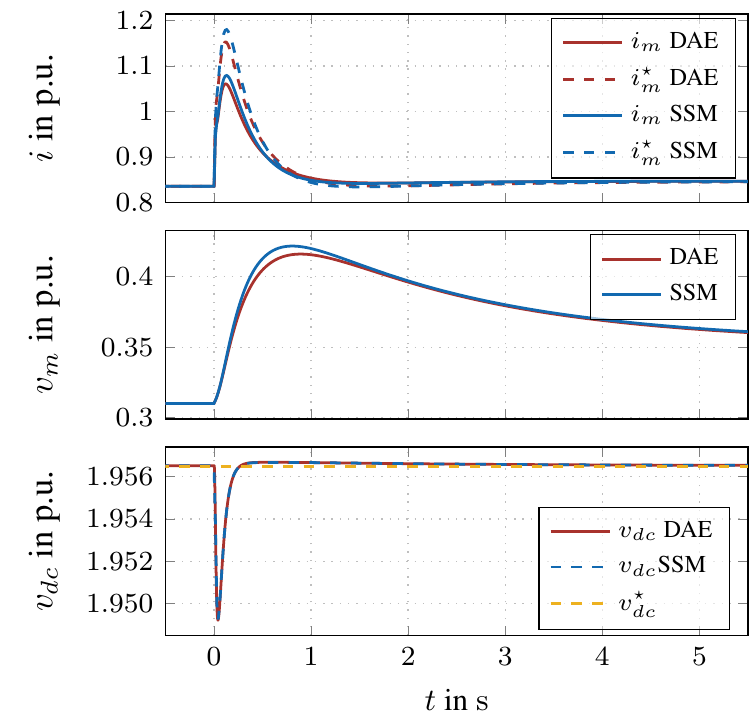}
    \caption{BLDC time-domain response to a step change in power system load.}
    \label{fig:BLDCTimeDomain}
\end{figure}
\begin{figure}[b!]
    \centering
    \includegraphics{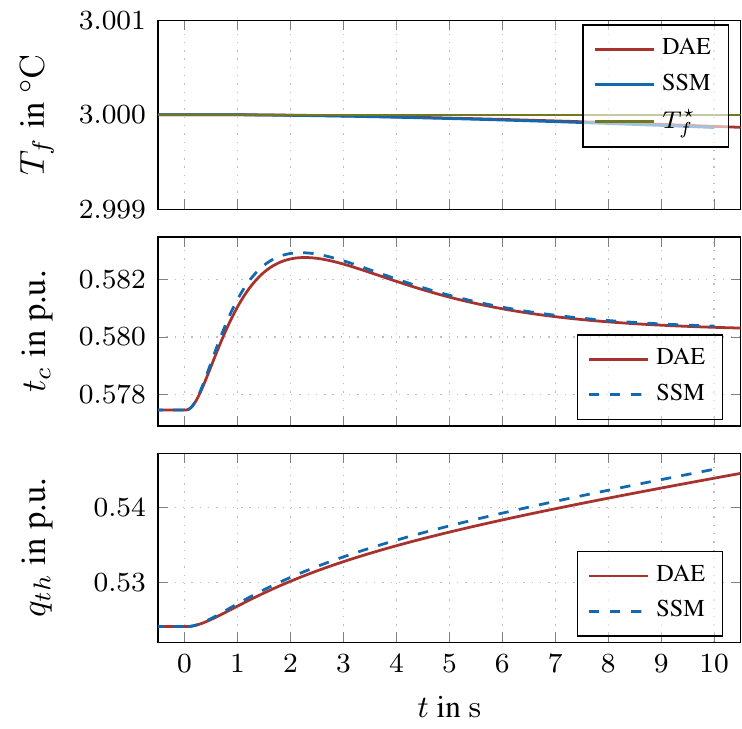}
    \caption{Time-domain response of the thermal system states and the compressor to a step change in power system load.}
    \label{fig:ThermalStates}
\end{figure}

The fast activation of VSDR is clearly showcased in Fig.~\ref{fig:GridResponse}, as the PLL accurately estimates the grid frequency and permits fast control of the reserve. The aggregated power closely follows the transients of the grid frequency and the reserve is activated almost instantaneously. 

Figure~\ref{fig:SingleDevice} presents the rotational speed reference, the resulting power reference and the terminal power of a single device. Moreover, it demonstrates the operation of the combined control objectives in speed control. While the contribution from temperature mismatch stays constant, the grid-responding input follows the PLL frequency estimate closely. Furthermore, the speed control error is minimised and the rotational speed of the BLDC follows its reference within $1\,\mathrm{s}$ after the disturbance, with the terminal power control error diminishing even faster.

The dynamics of the BLDC and VSD are displayed in Fig.~\ref{fig:BLDCTimeDomain}. Similar to the rotational speed in Fig.~\ref{fig:SingleDevice}, the motor current follows its reference closely but experiences a significant error during the transients. While the DC-link voltage stays nearly constant, the PWM successfully manipulates the modulation voltage to achieve the desired behaviour. 

Despite the significant change in active power consumption, the activation of the VSDR control only marginally affects the instantaneous thermal energy supplied to the refrigeration chamber, as indicated in Fig.~\ref{fig:ThermalStates}. Due to the different time delays, load torque and power react much faster than thermal heat. Hence, the refrigeration compartment temperature remains constant on this timescale and consumer comfort is preserved, as long as the grid frequency will return to its nominal value within an acceptable timeframe.

\begin{figure}[b]
\vspace{-5pt}
\centering
\includegraphics{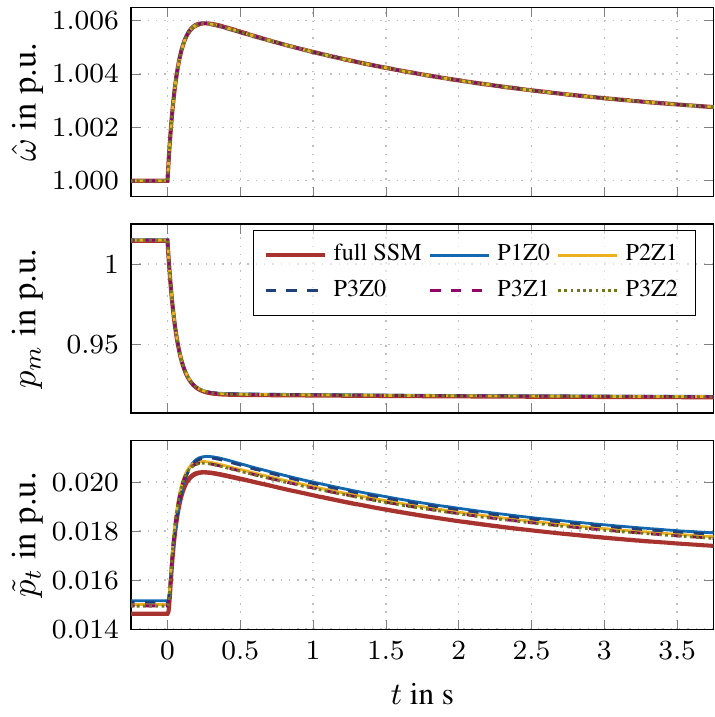}
    \caption{Grid-level response of the full-order and simplified SSMs to the $\SI{-0.1}{\pu}$ step change in load power.}
    \label{fig:TD_CL_gridLevel}
\end{figure}
\begin{figure}[b]
\centering
\includegraphics{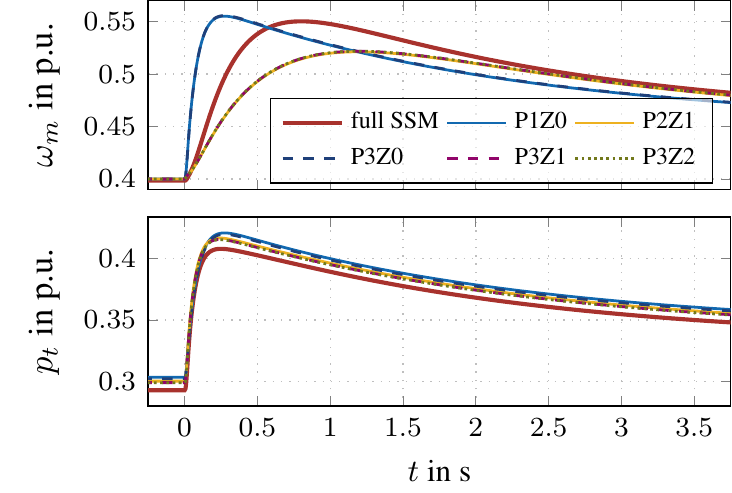}
    \caption{Single device response of the full-order and simplified SSMs to the $\SI{-0.1}{\pu}$ step change in load power.}
    \label{fig:TD_CL_SingleUnit}
\end{figure}

The results presented in the aforementioned Fig.~\ref{fig:GridResponse}-\ref{fig:ThermalStates} demonstrate the accuracy of the implemented small-signal model, i.e., moving from (\ref{eq:stateSpace}) to (\ref{eq:full_sys}). For the compressor torque and thermal power supplied to the compressor (see Fig.~\ref{fig:ThermalStates}) the small-signal model is slightly inaccurate and results in noticeable deviation from the full DAE results. These originate from the non-linear compressor model chosen in Section~\ref{sec:VSDRunit} and propagate to other states of the model.

\subsection{Model Comparison}
\begin{figure}[b]
    \centering
    \begin{subfigure}[t]{\linewidth}
       \includegraphics{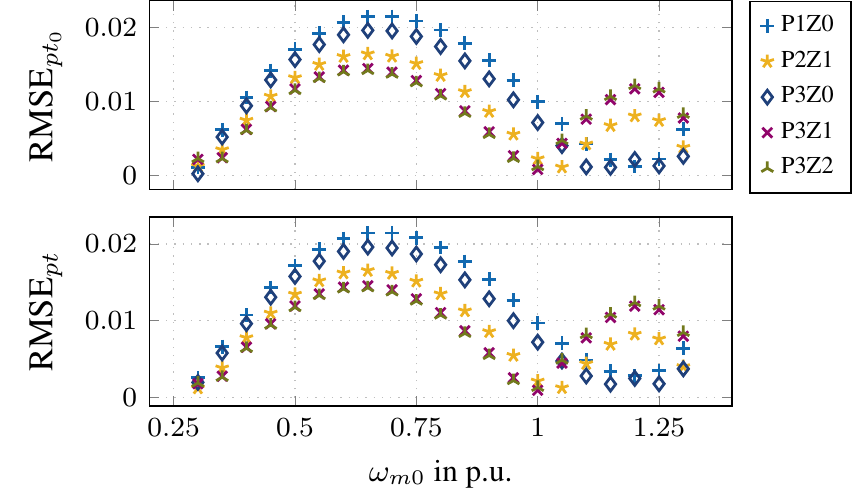}
       \caption{}
   \end{subfigure}\\[1em]
   \begin{subfigure}[t]{\linewidth}
        \includegraphics{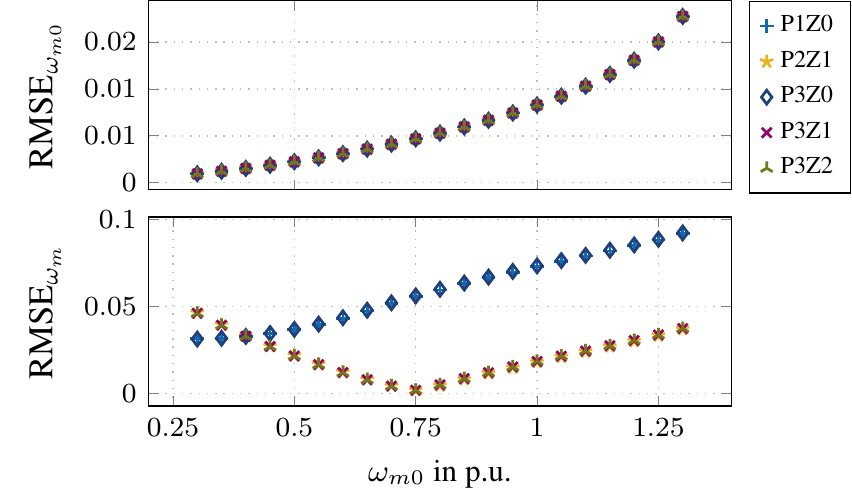}
        \caption{}
   \end{subfigure}
    \caption{Absolute RMSE of different variables during initialisation and transients for varying initial operating points: (a) terminal power; (b) rotational speed.}
    \label{fig:LinError}
\end{figure}
To test the limitations and validity of the proposed model reduction, the previous closed-loop simulations were repeated on the reduced-order models, with the results shown in Fig.~\ref{fig:TD_CL_gridLevel}. The model $P2Z0$ is excluded in the forthcoming analysis due to experiencing instability for the given operation point. All other simplified models slightly overestimate the single VSDR power right after the disturbance, but accurately capture the grid frequency response and mechanical power for the aggregation. Moreover, Fig.~\ref{fig:TD_CL_SingleUnit} reflects the time-domain response of a single device. Even though all models either underestimate or overestimate the response time of the rotational speed, all of them accurately capture the terminal power evolution.

In the following, we will analyse the effect of the initial operating point on the accuracy of the simplified models. Therefore, the initial rotational speed $\omega_{m0}$ is varied to span the entire BLDC operation range of \SIrange{0.3}{1.35}{\pu}, and the models are subjected to the same step change in load power. For comparison, we estimate the absolute root-mean square error~$\text{RMSE}$ pertaining to the initialisation of the simplified model (denoted by subscript 0), as well as the RSME during the transient response (denoted by subscript $t$). For a quantity $x$, the former is defined as
\begin{align}
    \text{RMSE}_{x0} &= \sqrt{\left( x_{0,\text{full}} - x_{0,\text{red}} \right)^2},
\end{align}
where $x_{0,\text{full}}$ and $x_{0,\text{red}}$ denote the initial magnitude of $x$ for the full- and reduced-order model, respectively. The RMSE of the transient response is considered for a time interval of $\SI{1}{s}$ after the disturbance, with a fixed time step resolution of $\SI{1}{\milli \second}$. Hence,
\begin{align}
    \text{RMSE}_{xt} &= \frac{1}{N} \sum _{k=0}^N \sqrt{\left( x_{k,\text{full}} - x_{k,\text{red}} \right)^2},
\end{align}
where $N$ is the number of included data points. 

The RMSE for the terminal power and rotational speed as well as their dependence on the initial operating point are depicted in Fig.~\ref{fig:LinError}. In general, all presented errors vary with the initial operating conditions. For the terminal power, both the initial and transient estimation errors show the same behaviour and peak for medium initial rotational speeds. As a result, the initialisation of the reduced-order model is the key reason for the mismatch between respective time-domain responses. On the other hand, the initialisation and transient errors for the rotational speed differ. While the initialisation error experiences the same increase for all models with increase in initial rotational speed, the transient errors can be classified into two groups: (i) the models without additional zeros that show a steady incline in transient error with initial speed; and (ii) the models with at least one additional zero which reach their minimum error for medium initial speeds. Here, the transient error dominates for all considered models.

In short, the RMSE in terminal power are expected to be kept below $\SI{0.02}{\pu}$ for the entire range of VSDR operation, while the RMSE during the transient response can reach up to \SI{0.1}{\pu} for the rotational speed. As a result, the reduced-order models are capable of accurately estimating the power output of a device aggregation, while the prediction of rotational speeds of the device is inferior. This might be of importance when computing activation margins or applying adaptive control schemes. 

\section{Stability Analysis} \label{sec5}
Besides accurately representing the time-domain performance, a suitable reduced-order model needs to capture the movement of system eigenvalues in order to be useful for tuning of an entire aggregation or performing small-signal stability studies on a larger scale. This section focuses in particular on modal analysis of the proposed models. After investigating the effect of the initial operating conditions, we conduct a bifurcation analysis to study the impact of frequency droop control and the grid equivalent on the stability margins.

\subsection{Initial Operating Conditions}
\begin{figure}[t]
\begin{minipage}[t]{\linewidth}
    \centering
    \includegraphics{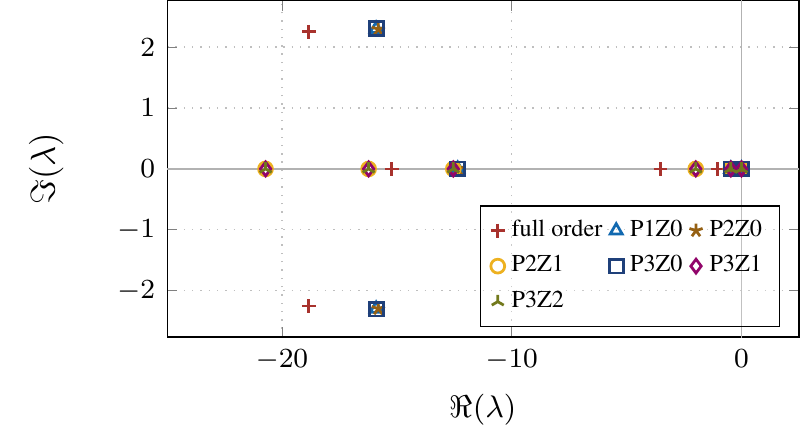}
    \caption{Root loci spectrum of interest for different SSMs.}
    \label{fig:eig_stable} 
    \vspace{1em}
    \scalebox{0.9}{\includegraphics{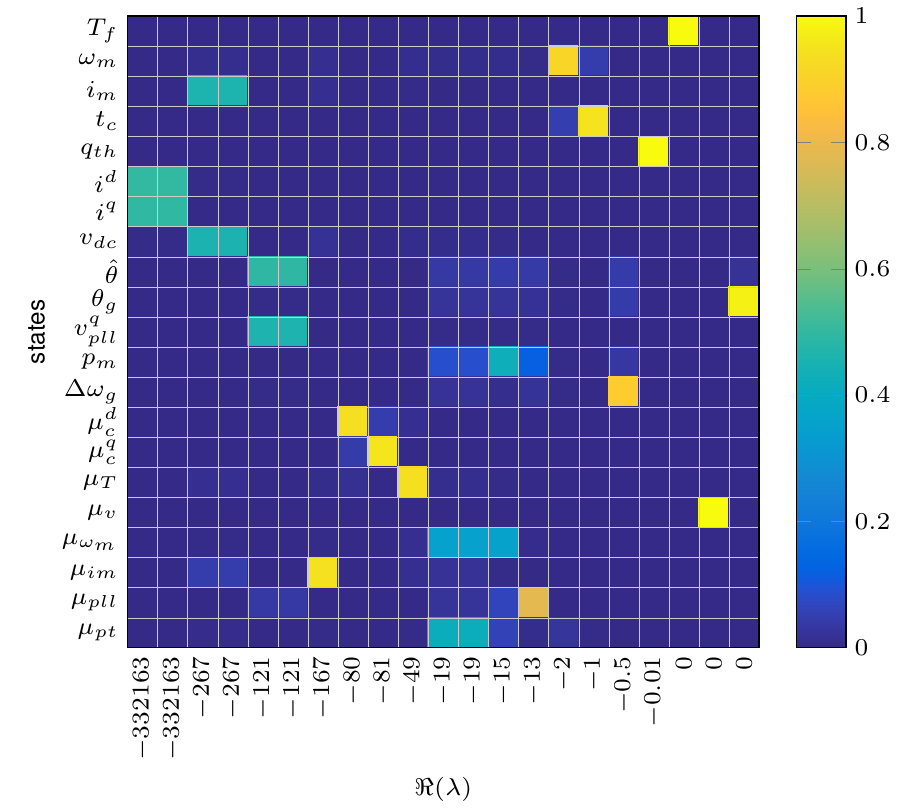}}
    \caption{Participation factors of the full-order SSM.}
    \label{fig:PF_Full}
    \vspace{1em}
    \includegraphics{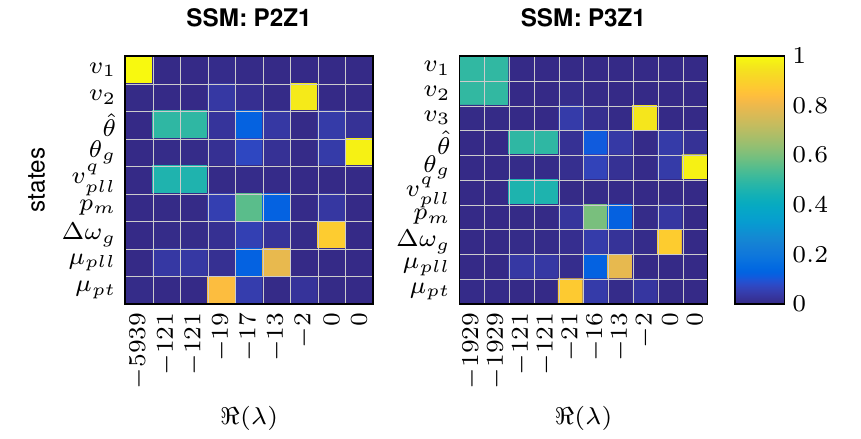}
    \caption{Participation factors of the reduced-order models $P2Z1$ and $P3Z2$.}
    \label{fig:PF_Red}
    \vspace{-20pt}
\end{minipage}
\end{figure}

The first stability study is conducted through eigenvalue analysis of the proposed small-signal models. The root loci spectrum of the most critical modes of the full- and reduced-order models is presented in Fig.~\ref{fig:eig_stable}. The positioning of the eigenvalues explains the similarities in time-domain performance of different simplified models. While the modes coincide for models without additional zeros, they also overlap for those having a zero. Consequently, since the modes of the former models are complex conjugate and close to the ones of the full-order system, these models are capable of capturing the slow oscillatory dynamics more accurately. 

Understandably, not all modes of the full-order system are represented by the simplified counterparts. The participation factor analysis (see Fig.~\ref{fig:PF_Full} and Fig.~\ref{fig:PF_Red} for full- and reduced-order models respectively) provides insights into which states might be relevant to each mode. We can observe that the modes close to the imaginary axis are not only related to the thermal model (i.e., the refrigeration temperature, the thermal energy and the compressor torque), but also to inner control loops (e.g., the DC voltage integrator state $\mu_v$) and the rotational speed ($\omega_m$). In addition, all models have modes close to or at the origin that originate from the grid equivalent. Interestingly, the simplified models preserve one mode with $\Re{\left(\lambda \right)} = -2$ that matches the mode corresponding to the rotational speed of the full-order equivalent.

It should be noted that the applied tuning of the full-order model provides a clear timescale separation of different cascaded control loops. Most of the integrator states clearly match one mode, i.e., the terminal current integrator, the motor current, the DC voltage integrator and the PLL states. On the other hand, the terminal power and rotational speed integrator states affect the same complex conjugate mode, indicating a link of these states.

\begin{figure}[t]
    \centering
    \includegraphics{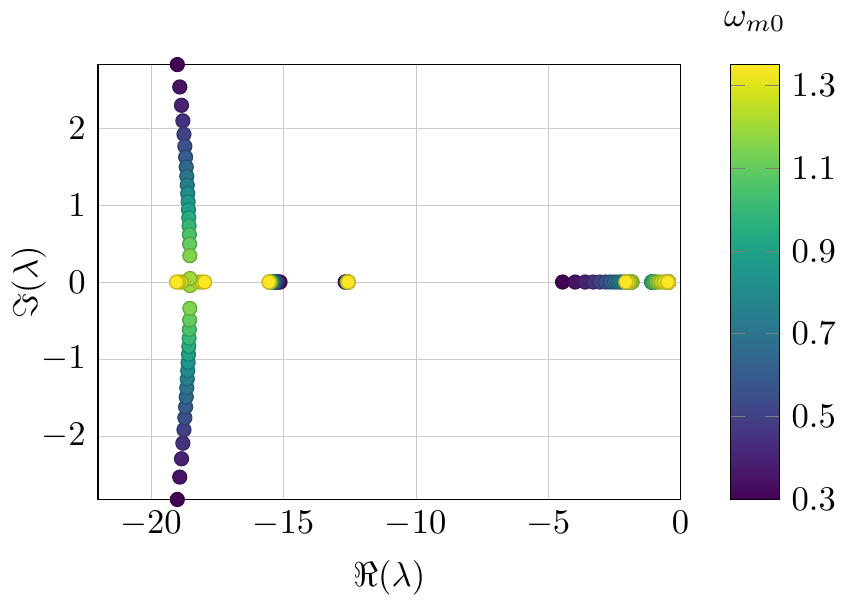}
    \caption{Movement of the most critical modes of the full order model with varying operation point.}
    \label{fig:EigMov_wm0}
\end{figure}
\begin{figure}[t]
    \centering
    \includegraphics{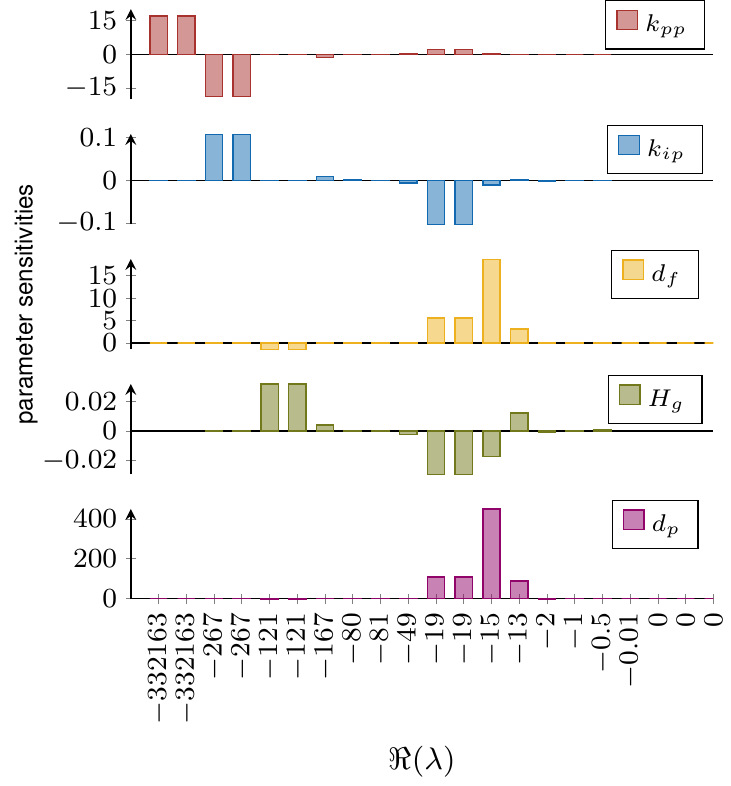}
    \caption{Selected parameter sensitivities of the detailed small-signal model.}
    \label{fig:parSens}
\end{figure}

\subsection{Effect of the Operating Point}
The effect of operating point on system performance is again revisited in this section. Fig.~\ref{fig:EigMov_wm0} displays the movement of the critical modes in the complex plane for the full-order model. While the displayed complex conjugate mode is increasingly damped with rising initial rotational speed, most non-oscillatory modes move significantly closer to the instability boundary. Hence, stability margins are reduced when the pre-fault rotational speed of the BLDC, resulting from a significant temperature deviation in the cooling chamber, is high. Nevertheless, such behaviour was not observed for simplified systems, as the modes remained unchanged. This is one of the main differences between the detailed and simplified models.

\subsection{Bifurcation Studies}
To identify which parameters might push the full-order model towards instability, we now investigate the impact of outer loop power control and the parameters of the grid equivalent via parameter sensitivities. After identifying the most critical parameters for the detailed model, we analyse if the same behaviour is preserved in the simplified models.

\begin{figure}
    \centering
    \includegraphics{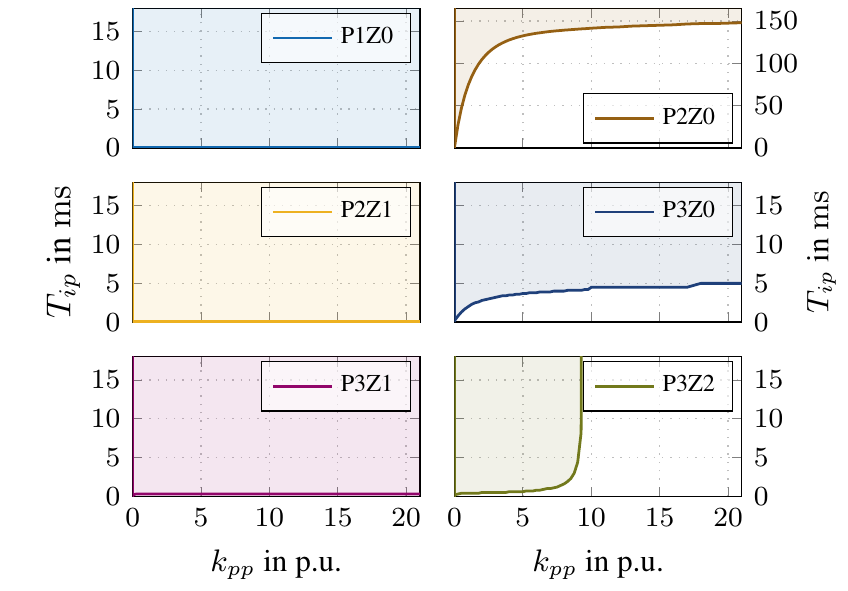}
    \caption{Stability map in $k_{pp}$-$T_{ip}$-plane for the simplified models. The shaded regions are stable.}
    \label{fig:StabMap}
\end{figure}
\begin{figure}
    \centering
    \includegraphics{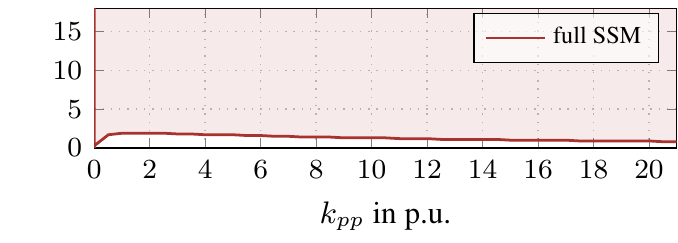}
    \caption{Stability map in $k_{pp}$-$T_{ip}$-plane of the full order model. The shaded region is stable.}
    \label{fig:StabMap_full}
\end{figure}
\begin{figure}
    \centering
    \includegraphics{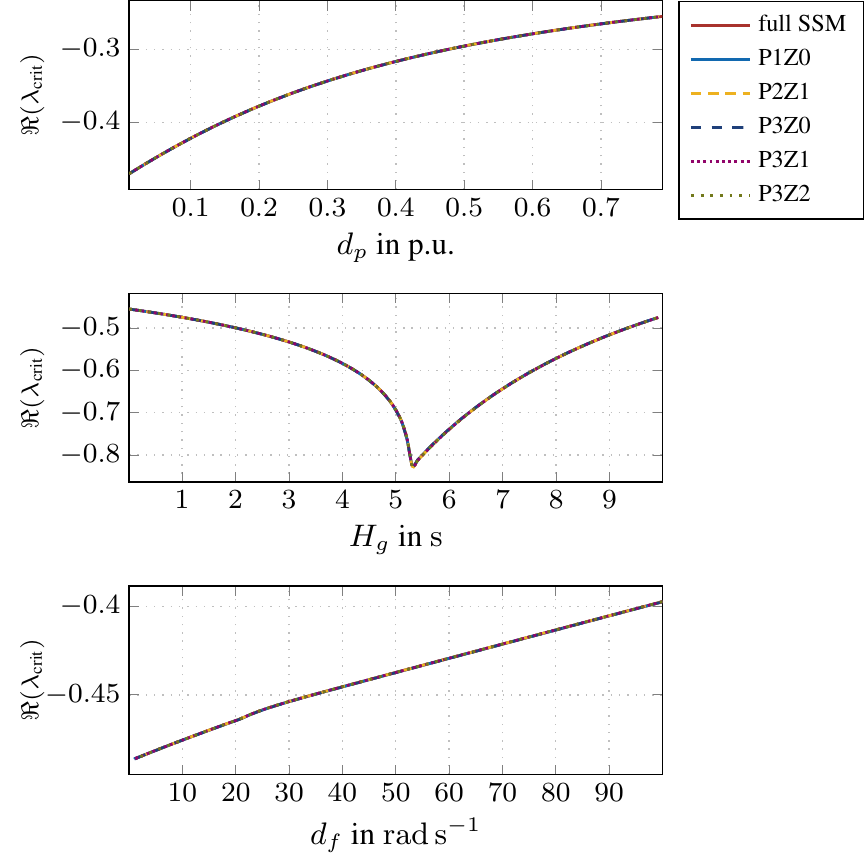}
    \caption{Movement of the critical eigenvalue with changing frequency droop gain, active power droop gain and system inertia for different models.}
    \label{fig:Sensitivities}
    \vspace{-10pt}
\end{figure}

The parameter sensitivities depicted in Fig.~\ref{fig:parSens} indicate that the terminal power control and the grid equivalent have little effect on the modes closest to the origin. Nonetheless, the frequency and active power droop appear to have an impact on few modes, primarily related to the integrator states of the terminal power, the rotational speed and the PLL, as well as the mechanical power of the grid equivalent. In contrast, the proportional gain of the power control mainly affects the fast electrical states, i.e., the terminal currents, DC-link voltage and motor current. Therefore, we only expect the frequency and active power droop gains to have a significant effect on the stability margins of the full-order model.

For the reduced-order models the situation differs. The stability maps provided in Fig.~\ref{fig:StabMap} imply a significant impact of power control tuning on small-signal stability. While the models $P1Z0$, $P2Z1$, $P3Z0$ and $P3Z1$ preserve stability for very small power control time constants $T_{ip}$ and high gains, similar to the full order model (see Fig.~\ref{fig:StabMap_full}), the other models experience different behaviour. More precisely, the P2Z0 model requires time constants above \SI{150}{\milli\second} to achieve stability, which is a threefold of the time constant the detailed system was designed for. At the same time, the $P3Z0$ model cannot be stabilised for proportional power control gains exceeding $10$. Thus, the latter two models should not be used for representing an aggregation, as they cannot replicate stability margins of the detailed system accurately and would be more conservative. 
Despite the discrepancies in power control tuning, all models behave similarly for different parameterisations of the grid equivalent and frequency droop control, as suggested by the results in Fig.~\ref{fig:Sensitivities}. Indeed, the critical mode movement is identical and stability is preserved, in particular for low-inertia scenarios. 

\section{Conclusion}
\label{sec6}
This paper proposes a droop control design of a single-phase-connected VSDR device and provides a detailed mathematical formulation of the underlying dynamic system. Moreover, a simple technique is presented for obtaining accurate reduced-order models that are suitable for large-scale power system studies. Comparing the full-order model to its reduced counterparts indicates significant overlap in both time-domain performance and small-signal stability margins. While most of the simplified models estimate the terminal power of a single device and the entire aggregation precisely, hence accurately replicating the frequency response of a system after being subjected to a disturbance in load, they fail to capture the rotational speed of the BLDC and thus the state and control margins of the BLDC device. Nonetheless, the time-domain simulation imply that some of the reduced-order models are useful for large-scale studies. Similar findings are established via the small-signal stability analysis, with most of the reduced-order models preserving stability margins. Moreover, the proposed simplified models support efficient parameter estimation in a real world implementation and provide acceptable accuracy for most of the test cases. 

Despite the overlap between simulations of the detailed and simplified models, employing the reduced systems should be carefully reconsidered when performing stability studies in low- or no-inertia power systems. This is particularly relevant when analysing interactions between converter-interfaced load and generation units, since the detailed model might still be necessary for capturing the system interactions on shorter timescales.

The closed-loop time domain simulations of the full- and reduced-order models for an aggregation of devices showcase not only smooth operation of the implemented system, but also confirm that VSDR devices are prominent candidates for providing FFR in low-inertia systems. The stability margins exhibit low sensitivity to power system inertia and are generally shown not to increase with higher inertia. 

Future work should study the dynamic interactions with other fast-reacting active units, such as converter-interfaced generation systems, and test the application of reduced order models in such cases. On the other hand, the wide time scales present in the proposed VSDR control scheme might result in cross couplings of the dynamics of synchronous machines and the electromagnetic transients in the network. Thus, the interactions between synchronous machines and the proposed VSDR reserve remain to be studied. For these considerations, the location of the VSDR systems and an appropriate modelling of the network is of importance. 

\vspace{-5pt}
\bibliographystyle{IEEEtran}
\bibliography{bibliography}

\begin{thebibliography}{10}
\providecommand{\url}[1]{#1}
\csname url@samestyle\endcsname
\providecommand{\newblock}{\relax}
\providecommand{\bibinfo}[2]{#2}
\providecommand{\BIBentrySTDinterwordspacing}{\spaceskip=0pt\relax}
\providecommand{\BIBentryALTinterwordstretchfactor}{4}
\providecommand{\BIBentryALTinterwordspacing}{\spaceskip=\fontdimen2\font plus
\BIBentryALTinterwordstretchfactor\fontdimen3\font minus
  \fontdimen4\font\relax}
\providecommand{\BIBforeignlanguage}[2]{{%
\expandafter\ifx\csname l@#1\endcsname\relax
\typeout{** WARNING: IEEEtran.bst: No hyphenation pattern has been}%
\typeout{** loaded for the language `#1'. Using the pattern for}%
\typeout{** the default language instead.}%
\else
\language=\csname l@#1\endcsname
\fi
#2}}
\providecommand{\BIBdecl}{\relax}
\BIBdecl

\bibitem{Milano18}
F.~Milano, F.~Dörfler, G.~Hug, D.~J. Hill, and G.~Verbic, ``{Foundations and
  Challenges of Low-Inertia Systems (Invited Paper)},'' \emph{2018 Power
  Systems Computation Conference (PSCC)}, 2018.

\bibitem{Fang19}
J.~Fang, H.~Li, Y.~Tang, and F.~Blaabjerg, ``{On the Inertia of Future
  More-Electronics Power Systems},'' \emph{IEEE Trans. Emerg. Sel. Topics Power
  Electron}, vol.~7, no.~4, pp. 2130--2146, 2019.

\bibitem{markovic_2019}
\BIBentryALTinterwordspacing
U.~Markovic, O.~Stanojev, E.~Vrettos, P.~Aristidou, and G.~Hug, ``Understanding
  stability of low-inertia systems,'' Feb 2019. [Online]. Available:
  \url{engrxiv.org/jwzrq}
\BIBentrySTDinterwordspacing

\bibitem{Fernandez20}
D.~Fernandez-Munoz, J.~I. Perez-Diaz, I.~Guisandez, M.~Chazarra, and
  A.~Fernandez-Espina, ``{Fast frequency control ancillary services: An
  international review},'' \emph{Renewable and Sustainable Energy Reviews},
  vol. 120, p. 109662, 2020.

\bibitem{ziras_primary_2015}
E.~Vrettos, C.~Ziras, and G.~Andersson, ``Fast and reliable primary frequency
  reserves from refrigerators with decentralized stochastic control,''
  \emph{IEEE Trans. Power Sys.}, vol.~32, no.~4, pp. 2924--2941, Jul. 2017.

\bibitem{kim_modeling_2015}
Y.-J. Kim, L.~K. Norford, and J.~L. Kirtley, ``Modeling and {Analysis} of a
  {Variable} {Speed} {Heat} {Pump} for {Frequency} {Regulation} {Through}
  {Direct} {Load} {Control},'' \emph{IEEE Trans. Power Syst.}, vol.~30, no.~1,
  pp. 397--408, Jan. 2015.

\bibitem{hui_equivalent_2019}
H.~Hui, Y.~Ding, and M.~Zheng, ``Equivalent {Modeling} of {Inverter} {Air}
  {Conditioners} for {Providing} {Frequency} {Regulation} {Service},''
  \emph{IEEE Trans. Ind. Electron.}, vol.~66, no.~2, pp. 1413--1423, Feb. 2019.

\bibitem{chakravorty_rapid_2017}
D.~Chakravorty, B.~Chaudhuri, and S.~Y.~R. Hui, ``Rapid {Frequency} {Response}
  {From} {Smart} {Loads} in {Great} {Britain} {Power} {System},'' \emph{IEEE
  Trans. Smart Grid}, vol.~8, no.~5, pp. 2160--2169, Sep. 2017.

\bibitem{Che19}
Y.~{Che}, J.~{Yang}, Y.~{Zhou}, Y.~{Zhao}, W.~{He}, and J.~{Wu}, ``Demand
  response from the control of aggregated inverter air conditioners,''
  \emph{IEEE Access}, vol.~7, pp. 88\,163--88\,173, 2019.

\bibitem{Azizipanah20}
R.~Azizipanah-Abarghooee and M.~Malekpour, ``{Smart Induction Motor Variable
  Frequency Drives for Primary Frequency Regulation},'' \emph{IEEE Trans.
  Energy Convers.}, vol.~35, no.~1, pp. 1--10, 2020.

\bibitem{Malekpour20}
M.~Malekpour, R.~Azizipanah-Abarghooee, F.~Teng, G.~Strbac, and V.~Terzija,
  ``{Fast Frequency Response From Smart Induction Motor Variable Speed
  Drives},'' \emph{IEEE Trans. Power Syst.}, vol.~35, no.~2, pp. 997--1008,
  2020.

\bibitem{Ibrahim20}
I.~Ibrahim, C.~O’Loughlin, and T.~O’Donnell, ``{Virtual Inertia Control of
  Variable Speed Heat Pumps for the Provision of Frequency Support},''
  \emph{Energies}, vol.~13, no.~8, p. 1863, 2020.

\bibitem{BFE}
{EnergieSchweiz: Bundesamt für Energie BFE}, ``{Energieetikette für Kühl-
  und Gefriergeräte},'' 2020.

\bibitem{FEA18}
{FEA Fachverband Elektroapparate für Haushalt und Gewerbe Schweiz},
  ``{Schweizer Verkaufsstatistik Geräte 2004-2016},'' Tech. Rep., 2018.

\bibitem{Gils14}
H.~C. Gils, ``{Assessment of the theoretical demand response potential in
  Europe},'' \emph{Energy}, vol.~67, pp. 1--18, 2014.

\bibitem{Grein06}
A.~Grein and M.~Pehnt, ``{Load management for refrigeration systems: Potentials
  and barriers},'' \emph{Energy Policy}, vol.~39, no.~9, pp. 5598--5608, 2011.

\bibitem{UCTE}
UCTE, ``{Appendix 1: Load-frequency control and performance},'' 2004.

\bibitem{koury_2001}
R.~Koury, L.~Machado, and K.~Ismail, ``\BIBforeignlanguage{en}{Numerical
  simulation of a variable speed refrigeration system},''
  \emph{\BIBforeignlanguage{en}{IJR}}, vol.~24, no.~2, pp. 192--200, Mar. 2001.

\bibitem{verhelst_2012}
C.~Verhelst, ``\BIBforeignlanguage{en}{Model {Predictive} {Control} of {Ground}
  {Coupled} {Heat} {Pump} {Systems} for {Office} {Buildings}},'' Ph.{D}.
  dissertation, KU Leuven, 2012.

\bibitem{li_feedforward_2009}
H.~Li, S.-K. Jeong, and S.-S. You, ``Feedforward control of capacity and
  superheat for a variable speed refrigeration system,'' \emph{Applied Thermal
  Engineering}, vol.~29, no.~5, pp. 1067--1074, Apr. 2009.

\bibitem{Secop_Data}
SECOP, \emph{Variable Speed Drive Compressor R600a}, May 2017, data sheet.

\bibitem{rasmussen_1997}
C.~B. Rasmussen and E.~Ritchie, ``Variable speed brushless {DC} motor drive for
  household refrigerator compressor,'' in \emph{8th International Conf. on
  Electrical Machines and Drives}, Jan 1997.

\bibitem{Samsung_2015}
\BIBentryALTinterwordspacing
Samsung, ``\BIBforeignlanguage{en}{How the {Digital} {Inverter} {Compressor}
  {Has} {Transformed} the {Modern} {Refrigerator}}.'' [Online]. Available:
  \url{www.news.samsung.com}
\BIBentrySTDinterwordspacing

\bibitem{kim_chap2_2017}
S.-H. Kim, ``Chapter 2 - {Control} of direct current motors,'' in
  \emph{Electric {Motor} {Control}}, S.-H. Kim, Ed.\hskip 1em plus 0.5em minus
  0.4em\relax Elsevier, Jan. 2017, pp. 39--93.

\bibitem{krishnan_permanent_2017}
R.~Krishnan, \emph{\BIBforeignlanguage{en}{Permanent {Magnet} {Synchronous} and
  {Brushless} {DC} {Motor} {Drives}}}, 1st~ed.\hskip 1em plus 0.5em minus
  0.4em\relax CRC Press, Dec. 2017.

\bibitem{kim_brushless_2017}
S.-H. Kim, ``\BIBforeignlanguage{en}{Chapter 10 - brushless direct current
  motors},'' in \emph{\BIBforeignlanguage{en}{Electric {Motor}
  {Control}}}.\hskip 1em plus 0.5em minus 0.4em\relax Elsevier, 2017, pp.
  389--416.

\bibitem{xia_permanent_2012}
C.-l. Xia, \emph{\BIBforeignlanguage{en}{Permanent magnet brushless {DC} motor
  drives and controls}}.\hskip 1em plus 0.5em minus 0.4em\relax Hoboken, NJ:
  Wiley, 2012.

\bibitem{hong-seok_song_advanced_2003}
{Hong-Seok Song}, R.~Keil, P.~Mutschler, J.~van~der Weem, and {Kwanghee Nam},
  ``Advanced control scheme for a single-phase {PWM} rectifier in traction
  applications,'' in \emph{38th {IAS} {Annual} {Meeting} on {Conf.} {Record} of
  the {Industry} {Applications} {Conf.}, 2003.}, vol.~3.\hskip 1em plus 0.5em
  minus 0.4em\relax Salt Lake City, UT, USA: IEEE, 2003.

\bibitem{bacha_power_2014}
S.~Bacha, I.~Munteanu, and A.~I. Bratcu, \emph{\BIBforeignlanguage{en}{Power
  {Electronic} {Converters} {Modeling} and {Control}}}, ser. Advanced
  {Textbooks} in {Control} and {Signal} {Processing}.\hskip 1em plus 0.5em
  minus 0.4em\relax London: Springer London, 2014.

\bibitem{hackl_equivalence_nodate}
C.~M. Hackl, ``\BIBforeignlanguage{en}{On the equivalence of
  proportional-integral and proportional-resonant controllers with
  anti-windup},'' p.~10.

\bibitem{yang_modeling_2018}
Y.~Yang and K.~Zhou, ``\BIBforeignlanguage{en}{Modeling and {Control} of
  {Single}-{Phase} {AC}/{DC} {Converters}},'' in
  \emph{\BIBforeignlanguage{en}{Control of {Power} {Electronic} {Converters}
  and {Systems}}}.\hskip 1em plus 0.5em minus 0.4em\relax Elsevier, 2018, pp.
  93--115.

\bibitem{ciobotaru_2006}
M.~Ciobotaru, R.~Teodorescu, and F.~Blaabjerg, ``A new single-phase {PLL}
  structure based on second order generalized integrator,'' in \emph{Proc. of
  IEEE PESC 2006}, Jun. 2006.

\bibitem{xiao_2017}
F.~Xiao, L.~Dong, L.~Li, and X.~Liao, ``A {Frequency}-{Fixed} {SOGI}-{Based}
  {PLL} for {Single}-{Phase} {Grid}-{Connected} {Converters},'' \emph{IEEE
  Trans. Power Electron.}, vol.~32, no.~3, pp. 1713--1719, Mar. 2017.

\bibitem{golestan_2013}
S.~Golestan, M.~Monfared, F.~D. Freijedo, and J.~M. Guerrero, ``Dynamics
  {Assessment} of {Advanced} {Single}-{Phase} {PLL} {Structures},'' \emph{IEEE
  Trans. Ind. Electron.}, vol.~60, no.~6, pp. 2167--2177, Jun. 2013.

\bibitem{rocabert_2012}
J.~Rocabert, A.~Luna, F.~Blaabjerg, and P.~Rodríguez,
  ``\BIBforeignlanguage{en}{Control of {Power} {Converters} in {AC}
  {Microgrids}},'' \emph{\BIBforeignlanguage{en}{{IEEE} Trans. Power
  Electron.}}, vol.~27, no.~11, pp. 4734--4749, Nov. 2012.

\bibitem{weckesser_2017}
T.~Weckesser and T.~{Van Cutsem}, ``Equivalent to represent inertial and
  primary frequency control effects of an external system,'' \emph{Transmission
  Distribution IET Generation}, vol.~11, pp. 3467--3474, 2017.

\bibitem{misyris_2019}
G.~S. Misyris, S.~Chatzivasileiadis, J.~A. Mermet~Guyennet, and J.~T.~G.
  Weckesser, ``\BIBforeignlanguage{en}{Grid {Supporting} {VSCs} in {Power}
  {Systems} with {Varying} {Inertia} and {Short}-{Circuit} {Capacity}},'' in
  \emph{\BIBforeignlanguage{en}{Proc. of {IEEE} {PES} {PowerTech} 2019}}, 2019.

\bibitem{Mahdavi20}
N.~Mahdavi and J.~H. Braslavsky, ``{Modelling and control of ensembles of
  variable-speed air conditioning loads for demand response},'' \emph{IEEE
  Trans. Smart Grid}, vol.~PP, no.~99, pp. 1--1, 2020.

\bibitem{markovic_stability_2018}
U.~Markovic, J.~Vorwerk, P.~Aristidou, and G.~Hug, ``Stability {Analysis} of
  {Converter} {Control} {Modes} in {Low}-{Inertia} {Power} {Systems},'' in
  \emph{{IEEE} {PES} {ISGT Europe} 2018}, Oct. 2018.

\bibitem{Kemmler18}
A.~Kemmler, T.~Spillmann, and S.~Koziel, ``{Der Energieverbrauch der Privaten
  Haushalte 2000 – 2017: Auswertung nach Verwendungszwecken und Ursachen der
  Veränderungen},'' Prognos AG, Tech. Rep., 2018.

\end{thebibliography}
\end{document}